\documentclass[a4paper,11pt]{article}
\usepackage{jheppub} 
\usepackage{amsmath,amssymb,array}
\usepackage{graphicx}
\usepackage{textcomp}
\usepackage[T1]{fontenc} 
\usepackage{color}
\usepackage{subfigure}
\usepackage{feynmp}
\usepackage{tabularray}
\usepackage{enumerate}
\usepackage{slashed}
\usepackage{arydshln} 
\usepackage{multirow} 
\usepackage{bm} 
\usepackage{hhline}
\usepackage{rotating}
\usepackage{hyperref}
\usepackage{booktabs}
\usepackage{xcolor}
\hypersetup{colorlinks=true,
	linkcolor=blue,
	filecolor=magenta,      
	urlcolor=blue}
\newcommand{\beqa}{\begin{eqnarray}}
\newcommand{\eeqa}{\end{eqnarray}}
\newcommand{\be}{\begin{equation}}
\newcommand{\ee}{\end{equation}}
\newcommand{\ba}{\begin{array}}
\newcommand{\ea}{\end{array}}

\title{\boldmath Flavour hierarchies from radiative corrections in latticed theory space}

\author[a]{Gurucharan Mohanta,}
\affiliation[a]{Asia Pacific Center for Theoretical Physics, Pohang 37673, Korea}
\author[b]{Ketan M. Patel}
\affiliation[b]{Theoretical Physics Division, Physical Research Laboratory,\\
Navarangpura, Ahmedabad-380009, India}

\emailAdd{gurucharan.mohanta@apctp.org}
\emailAdd{ketan.hep@gmail.com}

\abstract{It has recently been shown that when $N_f$ generations of chiral fermions are coupled in a specific manner to $N$ (with $N \geq 2N_f-1$) pairs of vectorlike fermions whose mass terms form a one-dimensional lattice-like structure in theory space, locality along the lattice ensures that only a single fermion generation acquires a mass at tree level. Radiative corrections can induce controlled departures from locality in the latticed space, thereby generating suppressed but non-vanishing masses for the remaining $N_f-1$ generations. In this work, we present an explicit implementation of this mechanism to address the flavour hierarchies of the Standard Model. After delineating the minimal extensions of the gauge, scalar, and Yukawa sectors required for feasible implementation of the mechanism, we demonstrate that the framework successfully reproduces the observed charged-fermion mass spectrum and quark mixing pattern. We analyse the new-physics effects arising from the extended sectors and confront them with existing constraints from direct, indirect searches and precision measurements. It is shown that a viable realisation of the mechanism allows the spectrum of vectorlike fermions and additional gauge boson to lie at scales as low as $\mathcal{O}(5)\,\mathrm{TeV}$ with the lightest states typically corresponding to top partners. This stands in sharp contrast to conventional radiative mass-generation scenarios, in which phenomenological constraints typically impose a lower bound on the new-physics scale of order a few hundred to several thousand TeV.}

\begin{document}
\maketitle
\flushbottom

\section{Introduction}
\label{sec:intro}
Seeking a more insightful understanding of the seemingly mysterious flavour structure of the Standard Model (SM) has been one of the strongest motivations for constructing theories beyond it. The ultimate goal is to formulate a more fundamental theory in which the origin of the three fermion generations is understood, and their various Yukawa couplings emerge as calculable parameters. While nothing has yet come close to achieving this, several ideas have been put forward to address different aspects of this puzzle; see, for example, \cite{Feruglio:2015jfa,Feruglio:2025ztj,Altmannshofer:2025rxc} for recent reviews. A noteworthy approach in this direction is based on the postulate that only the third-generation fermions acquire masses at leading order, while the remaining generations obtain comparatively suppressed masses through radiative corrections \cite{Weinberg:1972ws,Georgi:1972hy,Mohapatra:1974wk,Barr:1978rv,Wilczek:1978xi,Yanagida:1979gs,Barbieri:1980tz}. In such frameworks, the latter can become calculable parameters of the theory. This idea is further motivated by the observation that the inter-generation mass hierarchies are of the order of loop-suppression factors, lending qualitative support to this mechanism.

\begin{figure}[t!]
\centering
\subfigure{\includegraphics[width=0.95\textwidth]{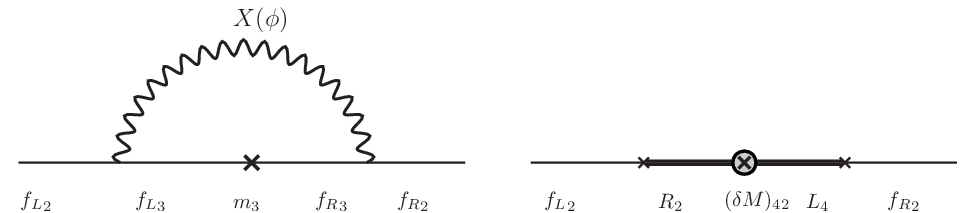}}
\caption{Diagrammatic representation of the next-to-leading-order origin of fermion masses in two approaches; see the text for details. On the left, the second-generation SM fermions ($f_{L2}$, $f_{R2}$) couple directly to the loop mediator(s). On the right, $f_{L2}$ and $f_{R2}$ couple indirectly through vectorlike fermions, with the relevant couplings between the latter arising only radiatively. The simple crosses denote a tree-level mass insertion, while the encircled represents the one induced radiatively.}
\label{fig1}
\end{figure}

At the quantitative level, the mechanism, however, faces two main challenges. First, it necessarily requires new fields beyond the standard model (BSM) that can induce the desired loop correction topology, see for example the left panel in Fig. \ref{fig1}. They can be scalars \cite{Balakrishna:1987qd,Balakrishna:1988ks,Balakrishna:1988xg,Balakrishna:1988bn,Babu:1988fn,Babu:1989tv,He:1989er,Rattazzi:1990wu,Berezhiani:1991ds,Berezhiani:1992bx,Berezhiani:1992pj,Arkani-Hamed:1996kxn,Barr:2007ma,Graham:2009gr,Dobrescu:2008sz,Crivellin:2011sj,CarcamoHernandez:2016pdu,Nomura:2016emz,Chiang:2021pma,Baker:2020vkh,Baker:2021yli,Yin:2021yqy,Chang:2022pue,Zhang:2023zrn,Bonilla:2023wok,Greljo:2023bix,Arbelaez:2024rbm,Greljo:2024zrj,Kuchimanchi:2024nkt} and/or gauge bosons \cite{HernandezGaleana:2004cm,Reig:2018ocz,Weinberg:2020zba,Jana:2021tlx,Mohanta:2022seo,Mohanta:2023soi,Mohanta:2024wcr,Jana:2024icm} with the latter being more economical due to the restrictive nature of the gauge interactions. Extension to Higgs and/or fermion sectors is also required to ensure the rank-one structure of the leading-order fermion mass matrices. The incorporation of new fields increases the number of parameters in the theory, and many of them remain incalculable in the viable models \cite{Mohanta:2022seo,Mohanta:2023soi,Mohanta:2024wcr,Mohanta:2024wmh,Jana:2024icm}.  Secondly, the BSM fields responsible for generating the desired radiative corrections also induce unsuppressed flavour-changing neutral current (FCNC), as can be seen from the diagram in Fig. \ref{fig1}. This typically pushes the new-physics scale to the multi-thousand-TeV range \cite{Mohanta:2022seo,Mohanta:2023soi,Mohanta:2024wcr,Mohanta:2024wmh} (or to about 200 TeV in a more optimised scenarios \cite{Mohanta:2025wiq}), rendering the framework inaccessible to direct experimental verification. The significant separation between the electroweak and new physics scales is considered an unwelcome feature also from the perspective of technical naturalness.

Recently, a qualitatively different mechanism for the radiative generation of fermion masses has been proposed in \cite{Patel:2025hol}. It uses a theory-space lattice-like structure \cite{Arkani-Hamed:2001nha,Arkani-Hamed:2001kyx,Hill:2000mu,Choi:2014rja,Kaplan:2015fuy} for the mass terms among $N$ pairs of vectorlike fermions, a construction qualitatively similar to that used in the models of flavour hierarchies based on clockwork \cite{Giudice:2016yja,Patel:2017pct,Alonso:2018bcg,vonGersdorff:2017iym,AbreudeSouza:2019ixc,Altmannshofer:2021qwx} or Anderson localisation \cite{Rothstein:2012hk,Craig:2017ppp}. These fermions have only onsite and off-site nearest-neighbour mass terms. It is shown in \cite{Patel:2025hol} that if $N_f$ pairs of chiral fermions are attached to this chain in a specific manner, only one of them acquires a mass, while $N_f - 1$ remain massless, provided $N \geq (2N_f - 1)$. This is a consequence of the strict locality of the lattice structure in theory space. Once the locality is broken in a mild and ordered way, the $N_f - 1$ generations of fermions can acquire masses. The desired non-locality can naturally arise through quantum corrections if the vectorlike fermions possess additional interaction(s). The latter can induce controlled and calculable non-local mass terms at next-to-leading order, providing non-zero but small masses for the otherwise massless fermions. The mechanism is therefore fundamentally different from other lattice theory-space models \cite{Giudice:2016yja,Patel:2017pct,Alonso:2018bcg,AbreudeSouza:2019ixc,Altmannshofer:2021qwx,Craig:2017ppp,Vempati:2025dow}, in which all fermion masses arise at leading order itself, and the inter-generational mass hierarchy is arranged through exponentially small zero-mode couplings.

The manner in which the radiative corrections induce masses for the lighter generations is also different from the previously considered approach based on the radiative mass generation mechanism. It is depicted in the right panel of Fig. \ref{fig1} for comparison. The noteworthy difference between the two is that the SM fermions need not directly couple to the new gauge boson(s) or scalar(s) that induce the radiative corrections. The corrections happen in the lattice theory space and involve only the vectorlike fermions directly. Therefore, large FCNCs can be avoided, and the new physics scale can be brought close to the electroweak scale, improving upon the aspects outlined above.

The present investigation aims to develop an explicit, minimal, and phenomenologically viable implementation of this mechanism within the SM to explain the observed charged fermion mass hierarchies. We show that the viable implementation of the mechanism requires at least five pairs of chain fermions in each sector and possibly the simplest extension in the scalar and gauge sectors. We derive the BSM effects that arise from these extensions and conduct a comprehensive study of the current phenomenological constraints on them, as determined by various relevant observables. It is shown that the fully realistic version of this mechanism can lie just one or two orders of magnitude above the electroweak scale, unlike the traditional radiative mass generation models.

The remainder of the paper is organised as follows. Section \ref{sec:mech} reviews the basic mechanism. Its implementation within the SM and the resulting implications are discussed in Section \ref{sec:SM}. The BSM implications are analysed in Section \ref{sec:bsm}. Section \ref{sec:benchmark} presents explicit numerical solutions followed by a comprehensive discussion on the associated phenomenological constraints in Section \ref{sec:pheno}. We also outline possible extensions of the model to account for neutrino masses in Section \ref{sec:nu_mass}. Finally, Section \ref{sec:summary} summarises our results and compares this framework with previous radiative mass-generation mechanisms. Two Appendices include some relevant and technical supplementary material.

\section{The Mechanism}
\label{sec:mech}
The underlying mechanism of generating the flavour hierarchies is based on the one proposed in \cite{Patel:2025hol}. Consider the $N_f$ generations of chiral fermions, $f_{L \alpha}$ and $f_{R \alpha}$, interacting with $N$ copies of vectorlike fermions in a specific way given by the following mass Lagrangian,
\beqa \label{Lm}
-{\cal L}_m &=& \sum_{\alpha=1}^{N_f}\, \left(\mu_\alpha\,\overline{f}_{L \alpha}  R_\alpha\,+\,\mu_\alpha^\prime\,\overline{L}_{(N+1-\alpha)}  f_{R \alpha} \right) + \sum_{i,j=1}^N\, M^{(0)}_{ij}\,\overline{L}_i R_j\,+\,{\rm h.c.}\,.\eeqa
Here, $L_i$ and $R_i$ denote the left- and right-chiral components of $i^{\rm th}$ vectorlike fermion. They are assumed to form a one-dimensional chain in the theory space with only the diagonal and nearest-neighbour mass terms characterised by,
\be \label{MFC}
M^{(0)}_{ij} = W\,\left( \delta_{ij} + t\, (\delta_{i+1,j} + \delta_{i,j+1})\right)\,.\ee
The above structure of $M^{(0)}_{ij}$ is similar to the Hamiltonian considered in the tight-binding model by Anderson \cite{Anderson:1958vr}. Unlike \cite{Patel:2025hol}, we do not introduce disorder in the on-site interactions and instead choose universal strengths for both the on-site and nearest-neighbour off-site interactions. This choice is made only to achieve simplicity, as the underlying mechanism does not depend on the presence of order or disorder in the couplings among chain fermions.

The structured interactions depicted in Eqs. (\ref{Lm},\ref{MFC}) can be arranged by invoking a global symmetry, $G = \Pi_{i=1}^N U(1)_i$. Under $G$, the vectorlike fermion at the $j^{\rm th}$ site has a charge $q^{(i)}_j = \delta_{ji}$. The off-site interaction between the $j^{\rm}$ and $(j+1)^{\rm th}$ sites can result from a set of spurions, charged as $\pm (\delta_{ji} - \delta_{j+1,i})$. Their non-vanishing values in the vacuum break $G$ to a single vectorial $U(1)$. The chiral fermions $f_{L \alpha}$ and $f_{R \alpha}$ can have charges $\delta_{\alpha,i}$ and $\delta_{\alpha, N+1-i}$, respectively, under the $G$. This arrangement of $N$ number of $U(1)$ factors is conceptually similar to the so-called ``quivers'' and ``moose'' considered in four-dimensional quantum field theories earlier in \cite{Arkani-Hamed:2001kyx,Georgi:1985hf,Douglas:1996sw}.

The universality of the on-site and off-site mass terms for the chain fermions allows one to analytically compute their spectrum. One finds
\be \label{M_diag} 
O^T M^{(0)} O = {\rm Diag.}\left(m^{(0)}_1,m^{(0)}_2,...~,m^{(0)}_{N}\right)\,,\ee
where
\be \label{m0_k} 
m^{(0)}_k = W \left(1-2t\, \cos \left(\frac{k \pi}{N+1}\right) \right)\,,\ee
are in general complex eigenvalues of $M^{(0)}$, and $O$ is a real orthogonal matrix with elements,
\be \label{O_jk} 
O_{jk} = (-1)^{j+1}\,\sqrt{\frac{2}{N+1}}\,\sin\left( \frac{j k \pi}{N+1}\right) \equiv v^{(k)}_j\,,\ee
which is also identified with $j^{\rm th}$ element of eigenvector $v^{(k)}$ of $M^{(0)}$ with eigenvalue $m^{(0)}_k$. It is noteworthy that $O$ does not depend on the strength of couplings among the chain fermions.

The chiral and vectorlike fermions can be combined, 
\beqa \label{calF}
{\cal F}_{L} &=& (f_{L1},...,f_{L N_f},L_1,...,L_N)^T\,, \nonumber \\
{\cal F}_{R} &=& (f_{R1},...,f_{R N_f},R_1,...,R_N)^T\,, \eeqa
to construct full $(N_f+N) \times (N_f+N)$ tree-level mass matrix ${\cal M}^{(0)}$ such that Eq. (\ref{Lm}) becomes
\be \label{Lm2}
-{\cal L}_m =  \overline{\cal F}_L\,{\cal M}^{(0)}\,{\cal F}_R\,+\,{\rm h.c.}\,,\ee
with 
\be \label{calM0}
{\cal M}^{(0)} = \left(\ba{cc} \left(0\right)_{N_f \times N_f} & \left(\mu \right)_{N_f \times N} \\ \left(\mu^\prime \right)_{N\times N_f}  & \left(M^{(0)}\right)_{N \times N}\ea\right)\,.\ee
Here, the subscript represents the dimension of the corresponding block matrix. The elements of non-square matrices $\mu$ and $\mu^\prime$ can be read from Eq. (\ref{Lm}) as
\be 
\mu = \left((\mu_{\alpha} \delta_{\alpha \beta})_{N_f \times N_f}\,~~(0)_{N_f \times (N-N_f)}\right)\,,~~\mu^\prime = \left(\ba{c} (0)_{(N-N_f)\times N_f} \\ (\mu^\prime_\alpha \delta_{\alpha,N-1+\beta})_{N_f \times N_f}\ea \right)\,.\ee

\begin{figure}[t!]
\centering
\subfigure{\includegraphics[width=0.55\textwidth]{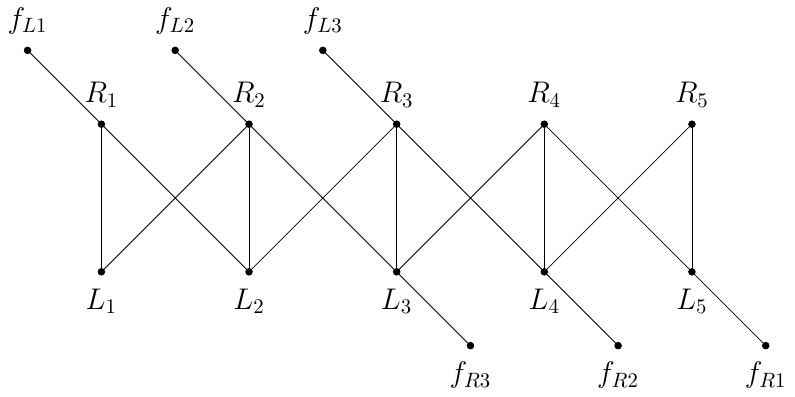}}
\caption{Topology of mass terms between the chiral and chain fermions as given by Eqs. (\ref{Lm},\ref{MFC}) for $N_f=3$ and $N=5$.}
\label{fig2}
\end{figure}

It is shown in \cite{Patel:2025hol} that ${\cal M}^{(0)}$ has rank $N+1$ for $N \geq (2N_f - 1)$, implying $(N_f - 1)$ massless modes. Therefore, to realise this mechanism in the context of the SM fermions so that the first- and second-generation remain massless at leading order, the minimal choice is $N = 5$. Accordingly, we set $N_f = 3$ and $N = 5$ in constructing an explicit flavour-hierarchy model based on this mechanism. The corresponding structure of mass terms is depicted in Fig. \ref{fig2}. The third-generation chiral fermions couple at the on-site through a vectorlike pair, while the remaining ones do not have localised couplings. We now discuss the implications of this particular arrangement. 

\subsection{Spectrum at the leading order}
With the aforementioned configuration of the latticed theory space, ${\cal M}^{(0)}$ in Eq. (\ref{calM}) becomes an $8 \times 8$ matrix with its blocks explicitly given by,
\be \label{tree_level_matrices}
\mu = \left( 
\ba{ccccc} \mu_1 &0 &0&0&0 \\ 0 &\mu_2 &0&0&0 \\ 0 &0 & \mu_3 &0&0\ea\right),~~ \mu^\prime = \left(\ba{ccc} 0 &0& 0 \\0 &0& 0 \\0 &0& \mu^\prime_3 \\0 &\mu^\prime_2 & 0 \\\mu^\prime_1 &0& 0 \ea \right),~~\text{and}~~ M^{(0)}= W\, \left( 
\ba{ccccc} 1 &t &0&0&0 \\ t &1 &t&0&0 \\0 &t &1&t&0 \\0 &0 &t&1&t \\ 0 &0 & 0 &t&1\ea\right)\,.\ee
The effective $3 \times 3$ light fermion mass matrix upon integrating out the vectorlike fermions, in the limit $|m_k^{(0)}| > |\mu_\alpha|, |\mu^\prime_\alpha|$, is then given by 
\be \label{meff_LO}
m^{(0)}_{\rm eff} \simeq - \mu\,{M^{(0)}}^{-1}\,\mu^\prime\,.\ee

The above matrix contains a massive and two massless states. The mass of the former can be estimated using
\beqa \label{tr_meff}
{\rm Tr}(m^{(0)}_{\rm eff}) &=& - \sum_{\alpha=1}^3\,\sum_{k=1}^5 \frac{\mu_\alpha \mu^\prime_\alpha}{m^{(0)}_k} v^{(k)}_{\alpha} v^{(k)}_{6-\alpha} = - \frac{(1-t^2)^2 \mu_3 \mu_3^\prime + t^2 \mu_2 \mu_2^\prime + t^4\, \mu_1 \mu_1^\prime}{(1-4t^2+3t^4)\, W}\,,\eeqa
which turns out to be $\sim\,{\cal O}(\mu_\alpha \mu^\prime_\alpha/W)$. This is the usual seesaw-induced mass, which is suppressed by the mass scale of the vectorlike fermions. Its magnitude is more or less independent\footnote{Note that $t=1$ leads to a massless vectorlike fermion in which case the seesaw limit fails.} of the strength of the off-site coupling.

The two massless states are linear combinations of five out of eight states given in ${\cal F}_{L,R}$. They are $\{f_{L 1},f_{L 2},f_{L 3},L_1,L_2\}$ for left-chiral and $\{f_{R 1}, f_{R 2}, f_{R 3}, R_4, R_5\}$ for the right-chiral components of the massless states \cite{Patel:2025hol}. The specific attachment of chiral fermions to the chain, together with the strict locality of lattice fermions, forbids any direct coupling among these states, thereby ensuring the masslessness of the two generations. Deviation from this locality can therefore give rise to small but non-vanishing masses for the otherwise massless states. One way to introduce such non-locality in a controlled and deterministic manner is to consider the radiative corrections as discussed in \cite{Patel:2025hol}. The non-local terms are loop-suppressed and typically ordered in a specific manner, leading to not only relatively suppressed masses for the massless states but also with some hierarchy among them.

\subsection{Radiative corrections}
The simplest way to introduce deterministic non-locality among the lattice fermions is to switch on gauge interactions among them. We consider a local $U(1)_X$ under which only the vectorlike fermions are charged with universal charges. Explicitly,
\be \label{L:gauge} 
-{\cal L}_X  = g_X q \left( \overline{L}_i \gamma^\mu L_i \,+ \,  \overline{R}_i \gamma^\mu R_i  \right)\, X_\mu \,.\ee
The mass terms $M^{(0)}_{ij}$ respect this symmetry while $\mu_\alpha$ and $\mu^\prime_\alpha$ can arise from the spontaneous breaking of it. Since the global symmetry group $G$ mentioned earlier is already broken by the nearest-neighbour off-diagonal terms in $M^{(0)}_{ij}$, the self-energy corrections induced by the new gauge interaction need not maintain the strict locality in $M^{(0)}$. 

An explicit evaluation of 1-loop self-energy corrections \cite{Patel:2025hol} lead to the shift in the $M^{(0)}_{ij}$ by,
\be \label{dM1}
\delta M^{(0)}_{ij} = \frac{g_X^2q^2}{4 \pi^2}\,\sum_{k=1}^5\, O_{ik}\,O_{jk}\,m^{(0)}_k\,B_0 \left[M_X^2,|m^{(0)}_k|^2 \right]\,,\ee
where the loop function $B_0$ is defined as 
\be \label{B0}
B_0 \left[a^2,b^2 \right] = \Delta_\epsilon + 1 - \frac{a^2 \ln \frac{a^2}{\mu_R^2} - b^2 \ln \frac{b^2}{\mu_R^2}}{a^2 - b^2}\,,\ee
with $\mu_R$ as  renormalisation scale, and 
\be \label{Delta_ep}
\Delta_\epsilon = \frac{2}{\epsilon} - \gamma_E + \ln 4\pi\,,\ee
parametrising the divergent part of loop-integration in $\overline{\rm MS}$ scheme \cite{Passarino:1978jh}. It is straightforward to verify from Eqs. (\ref{dM1},\ref{M_diag}) that the divergent part of the corrections is proportional to $M^{(0)}_{ij}$ itself and can therefore be renormalised. 

Denoting $B_0 \left[a^2,b^2 \right] - \Delta_\epsilon \equiv b_0 \left[a^2,b^2 \right]$ and using Eq. (\ref{O_jk}), the finite mass shift can be written as
\be \label{dM}
\delta M^{(0)}_{ij} = \frac{g_X^2q^2}{12\pi^2}\, (-1)^{i+j}\, \sum_{k=1}^5 \sin \left(\frac{i k \pi}{6}\right)\sin\left(\frac{j k \pi}{6}\right)\, m^{(0)}_k\,b_0 \left[M_X^2,|m^{(0)}_k|^2 \right]\,.\ee
The above not only corrects the non-vanishing $M^{(0)}_{ij}$ but also induces non-zero and finite non-local mass terms. The size of these terms depends on the hierarchy between $|m^{(0)}_k|$ and $M_X$. For $|m^{(0)}_k| \ll M_X$, the above can be approximated to,
\beqa \label{dM_l1}
\delta M^{(0)}_{ij} &\approx & \frac{g_X^2q^2}{4\pi^2} M^{(0)}_{ij} \left(1- \ln\frac{M_X^2}{\mu_R^2}\right) \nonumber \\
 &+& \frac{g_X^2q^2}{12\pi^2}\, (-1)^{i+j}\, \sum_{k=1}^5 \sin \left(\frac{i k \pi}{6}\right)\sin\left(\frac{j k \pi}{6}\right)\, m^{(0)}_k\, \frac{|m^{(0)}_k|^2}{M_X^2} \ln\frac{|m^{(0)}_k|^2}{M_X^2}\,,\eeqa
at the leading order in $|m^{(0)}_k|^2/M_X^2$. Similarly, for $|m^{(0)}_k| \gg M_X$, one finds
\beqa \label{dM_l2}
\delta M^{(0)}_{ij} &\approx & \frac{g_X^2q^2}{4\pi^2} M^{(0)}_{ij} \left(1- \ln\frac{M_X^2}{\mu_R^2}\right) \nonumber\\
 & - & \frac{g_X^2 q^2}{12\pi^2}\, (-1)^{i+j}\, \sum_{k=1}^5 \sin \left(\frac{i k \pi}{6}\right)\sin\left(\frac{j k \pi}{6}\right)\, m^{(0)}_k\, \ln\frac{|m^{(0)}_k|^2}{M_X^2}\,,\eeqa
at the leading order. The terms in the second line in Eqs. (\ref{dM_l1},\ref{dM_l2}) give rise to deviation from the locality in lattice fermion interactions. 

Since $f_{L \alpha}$, $f_{R \alpha}$ are not charged under the $U(1)_X$, they do not have direct couplings with the $X$-boson. Hence, the mass terms in $\mu$ and $\mu^\prime$ do not receive corrections like the above at the 1-loop.  With this, the 1-loop corrected full $8 \times 8$ mass matrix can be written as
\be \label{calM}
{\cal M} = \left(\ba{cc} \left(0\right)_{3 \times 3} & (\mu)_{3\times5}
\\ (\mu^\prime)_{5\times3}
& \left(M\right)_{5 \times 5}\ea\right)\,,\ee
with 
\be\label{M:1loop} 
M = M^{(0)} + \delta M^{(0)}\,.\ee

The diagonalisation of ${\cal M} $ can be carried out using unitary transformations, ${\cal F}_{L,R} = {\cal U}_{L,R}\, {\cal F}^\prime_{L,R}$, such that 
\be \label{88:diag} 
{\cal U}_L^\dagger \, {\cal M}\, {\cal U}_R = {\cal D}\,,\ee
and ${\cal F}^\prime_{L,R}$ are fields in the physical basis. In the limit of heavy vectorlike fermions, ${\cal U}_{L,R}$ can be further simplified to 
\be \label{88U_block}
{\cal U}_{L,R} =\left(\ba{cc} \left(\mathbb{I}_3-\frac{1}{2} \rho_{L,R} \rho_{L,R}^\dagger\right)\,u_{L,R} & -\rho_{L,R}\,U_{L,R} \\ \rho_{L,R}^\dagger\, u_{L,R} & \left(\mathbb{I}_5-\frac{1}{2} \rho_{L,R}^\dagger \rho_{L,R}\right)\, U_{L,R}\ea\right)\,+\,{\cal O}(\rho_{L,R}^3)\,.\ee
Here, $\rho_L = -\mu M^{-1}$ and $\rho_R^\dagger = -M^{-1} \mu^\prime$ are $3 \times 5$ and $5 \times 3$ matrices, respectively. $U_{L,R}$ are $5 \times 5$ matrices that diagonalize $M$ given in Eq. (\ref{M:1loop}). Explicitly,
\be \label{M1_diag}
U_L^\dagger\,M\,U_R = {\rm Diag.} \left(m_1,...,m_5 \right)\,,\ee
and $m_k$ are 1-loop corrected masses of the vectorlike fermions. 

The simple structures of $M^{(0)}$ and $\delta M^{(0)}$ allows one to express analytically the one-loop corrected spectrum as
\be \label{m_k}
m_k = \left| m^{(0)}_k \right |  \left(1 + \frac{g_X^2q^2}{4 \pi^2}\,b_0 \left[M_X^2,|m^{(0)}_k|^2 \right] \right)\,,\ee
and
\be \label{UR_UL}
(U_R)_{jk} = (U_L^*)_{jk} = O_{jk} \, e^{-\frac{i}{2}{\rm arg}\left[m^{(0)}_k\right]}\,\ee
with $m^{(0)}_k$ and $O_{jk}$ as given in Eqs. (\ref{m0_k},\ref{O_jk}) with $N=5$, respectively. 

Finally, the unitary matrices $u_{L,R}$ diagonalises the $3 \times 3$ effective mass matrix,
\be \label{m1_eff} 
m_{\rm eff} \simeq -\mu\, M^{-1}\, \mu^{\prime}\,. \ee
Unlike the $m^{(0)}_{\rm eff}$, this is not a rank-deficient matrix and all three generations can obtain non-vanishing masses. Indeed, for $\delta M^{(0)} \ll M^{(0)}$, the above can be approximated as
\be \label{m1_eff_1} 
m_{\rm eff} \approx - \mu {M^{(0)}}^{-1} \mu^{\prime} + \mu {M^{(0)}}^{-1} \delta M^{(0)} {M^{(0)}}^{-1} \mu^{\prime} \,, \ee
implying explicitly the ${\cal O} ((M^{(0)})^{-1} \delta M^{(0)})$ suppressed masses for the lighter generations in comparison to the leading order result. The $3\times 3$ unitary matrices $u_{L,R}$ in Eq. (\ref{88U_block}) diagonalizes $m_{\rm eff}$ such that
\be \label{m1eff_diag}
u_L^\dagger\,m_{\rm eff}\,u_R = {\rm Diag.} \left(m_{f_1}, m_{f_2}, m_{f_3} \right)\,,\ee
and $m_{f\alpha}$ are physical masses of the lighter three generations.

\begin{figure}[t!]
\centering
\subfigure{\includegraphics[width=0.45\textwidth]{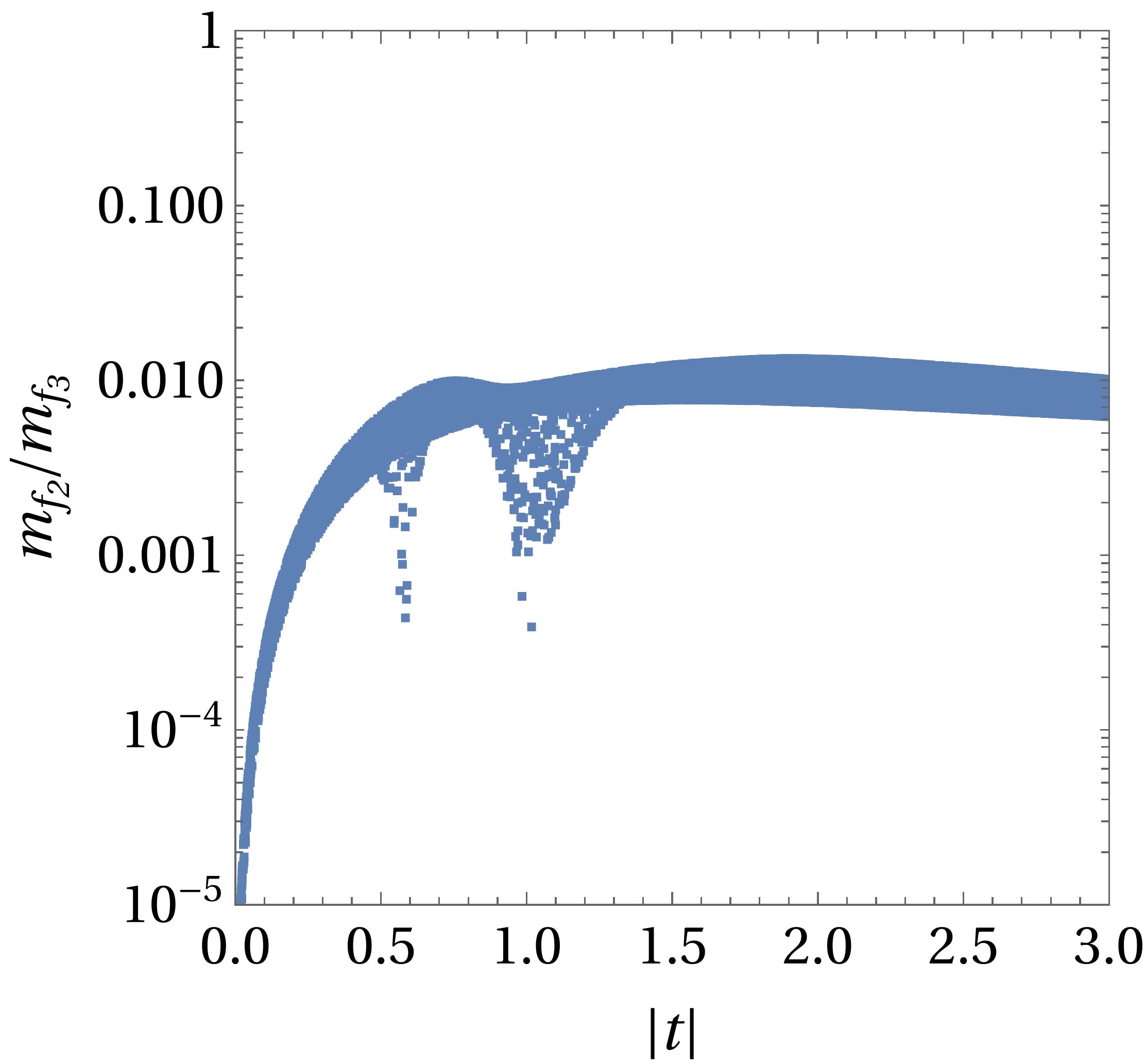}}\hspace*{0.1cm}
\subfigure{\includegraphics[width=0.45\textwidth]{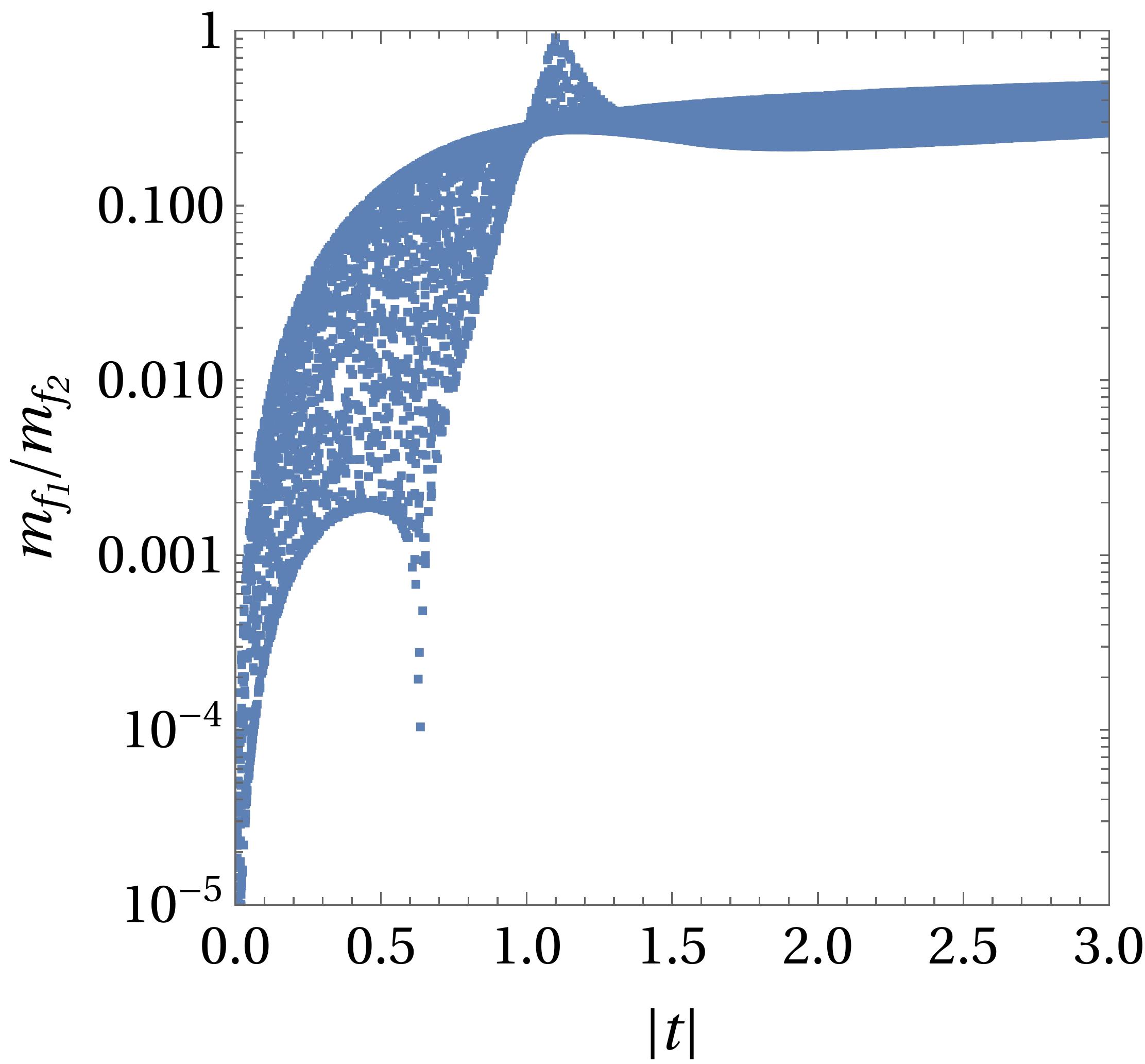}}
\caption{The mass ratios computed from Eqs. (\ref{m1_eff},\ref{m1eff_diag}) as a function of $t$ for $M_X = 10^4$ GeV, $\mu_\alpha = 174$ GeV, $\mu_\alpha = M_X/\sqrt{2}$ and $W = 2 M_X$.}
\label{fig3}
\end{figure}
It is a noteworthy aspect of the underlying mechanism that $m_{f_{1,2}}$ are not only suppressed due to their next-to-leading order origin in the theory, but a hierarchy among them is also arranged through the latticed structure \cite{Patel:2025hol}. We demonstrate this in Fig. \ref{fig3} in which the mass ratios $m_{f_2}/m_{f_3}$ and $m_{f_1}/m_{f_2}$ are computed for several values of complex $t$ and universal $\mu_\alpha$, $\mu^\prime_\alpha$.  As $|t| \to 0$, the fermions in the chain have no off-site mass terms, and they remain vanishing even at the next-to-leading order in the perturbation theory, leading to $m_{f_{1,2}} = 0$. For sizeable $|t|$, one finds $m_{f_2}/m_{f_3} \sim {\cal O}\left(\frac{1}{4 \pi^2}\right)$, as anticipated from Eq. (\ref{m1_eff_1}), almost independent of any specific value of $t$.  Furthermore, $0.5 \lesssim |t| \lesssim 1$ also lead to desired hierarchy in $m_{f_1}/m_{f_2}$ as it can be seen from Fig. \ref{fig3}.  As $|t| > 1$, the localisation becomes feeble, leading to less hierarchy among   $m_{f_1}$ and $m_{f_2}$ \cite{Patel:2025hol}. Also note that $t = \pm 1/\sqrt{3}$ or $\pm 1$ leads to an almost massless vectorlike fermion state, see Eq. (\ref{m0_k}), which is to be avoided in any case for phenomenological reasons.

\subsection{Distinction from other theory-space lattice models}
\label{subsec:distinction}
Before we discuss an implementation of the above mechanism on the SM fermions, let's point out some distinctive features with respect to the other models based on similar constructions. 

Small flavour hierarchies can be generated dynamically through localisation mechanisms such as clockwork \cite{Giudice:2016yja} and Anderson localisation \cite{Craig:2017ppp}, without introducing small fundamental parameters in four-dimensional quantum field theories. In the clockwork constructions, a chain of fields with structured nearest-neighbour couplings gives rise to a light zero mode whose wavefunction is exponentially localised in theory space. When the Higgs or Yukawa interactions are localised at a fixed site, fermions residing at different sites couple with exponentially varying strengths, leading to hierarchical effective Yukawa couplings from underlying $\mathcal{O}(1)$ parameters. The length of the chain (or the number of so-called clockwork gears) can be different for different generations of a given fermion type \cite{Giudice:2016yja,Alonso:2018bcg} or can be the same for all generations \cite{Patel:2017pct,vonGersdorff:2017iym}, depending on the specific implementation of the mechanism. In contrast, the models based on Anderson localisation \cite{Craig:2017ppp,Vempati:2025dow} rely on disorder, such as random on-site or off-site mass terms among the chain fermions, which induces exponential localisation of fermion wavefunctions at random positions. The resulting Yukawa couplings are determined by wavefunction overlaps and naturally exhibit a statistical hierarchy, with small values occurring with high probability.

In both cases discussed above, the flavour hierarchies originate from exponential suppression due to localisation of the wavefunction, while all fermion masses arise already at tree level. In contrast, the mechanism discussed in this paper gives rise to only one massive mode at tree level. The existence of $N_f - 1$ strictly massless modes follows from the locality of interactions in the one-dimensional chain in theory space. These modes acquire small, and in principle calculable, masses once non-locality is generated in a controlled and deterministic manner, which can be realised through radiative corrections, as discussed here. In this way, the mechanism does not require separate chains for each generation. In fact, it stems essentially from multiple copies of the field attached to a single chain. Moreover, while the small non-locality can be optional in the clockwork and Anderson model, it is essential in the present case for the realistic flavour spectrum.

\section{Standard Model Embedding}
\label{sec:SM}
The mechanism discussed above can be readily implemented for the SM fermions by extending it to include their vectorlike partners. The field content is displayed in Table \ref{tab:field} along with their gauge charges.
\begin{table}[!h]
\begin{center}
\begin{tabular}{cccc}
\toprule
~~Fields~~ & ~~Multiplicity~~  &~~ $G_{\rm SM}$~~&~~ $U(1)_X$\\
\midrule
$q_{L\alpha}=\left(u_{L \alpha}\,\,d_{L \alpha} \right)^T$ & $3$ & $ \left(3,2,\frac{1}{6} \right)$ &  $ 0$ \\
$u_{R\alpha}$ & $3$ & $\left(3,1,\frac{2}{3} \right) $& $ 0$ \\
$d_{R\alpha}$ & $3$ & $\left(3,1,-\frac{1}{3} \right) $& $ 0$ \\
$l_{L\alpha}=\left(\nu_{L \alpha}\,\,e_{L \alpha}\right)^T$ & $3$ & $ \left(1,2,-\frac{1}{2} \right)$& $ 0$ \\
$e_{R\alpha}$ & $3$ & $\left(1,1,-{1}\right)$ & $ 0$ \\
$H$ & $1$ &  $\left(1,2,\frac{1}{2} \right)$  & $ 1$ \\
\midrule
$U_{Li,Ri}$ & $5$ & $\left(3,1,\frac{2}{3} \right)$ & $1$\\
$D_{Li,Ri}$ & $5$ & $\left(3,1,-\frac{1}{3} \right)$ & $-1$\\
$E_{Li,Ri}$ & $5$ & $\left(1,1,-1 \right)$ & $-1$\\
$S$ & $1$ & $\left(1,1,0 \right) $ &  $-1$ \\
\bottomrule
\end{tabular}
\end{center}
\caption{The matter field content of the model along with their multiplicity and charges under $G_{\rm SM}=SU(3)_C \times SU(2)_L \times U(1)_Y$ and new $U(1)_X$.}
\label{tab:field}
\end{table}
The vectorlike fermions we introduce are singlets under the $SU(2)_L$. An additional singlet scalar $S$ is also introduced to break the new abelian gauge symmetry spontaneously as well as to enable gauge-invariant interactions between the chiral and vectorlike fields. In the following, we discuss in detail the scalar, gauge and Yukawa sectors of the model. 

\subsection{Scalar sector}
\label{subsec:scalar}
This is the simplest sector of the extension. The most general gauge-invariant renormalisable scalar potential of the theory can be written as
\be \label{V}
V = -\frac{\mu_H^2}{2} H^\dagger H + \frac{\lambda_H}{4} (H^\dagger H)^2 -\frac{\mu_S^2}{2} |S|^2 + \frac{\lambda_S}{4} |S|^4 + \frac{\lambda_{HS}}{2} H^\dagger H |S|^2\,,\ee
where all the parameters are real and positive. The vacuum expectation values (VEVs),
\be \label{VEVs} 
\langle  H \rangle = \left( \ba{c} 0\\ \frac{v}{\sqrt{2}}\\ \ea \right) \, ,\indent \langle S\rangle = \frac{v_{S}}{ \sqrt{2}} \,,\ee
with $v=246\, {\rm GeV}$ break the electroweak symmetry and $U(1)_X$. Minimisation of the potential with respect to these VEVs leads to
\be \label{min_cond}
2\mu_H^2 = \lambda_H\, v^2 + \lambda_{HS}\, v_S^2\,,\indent 2 \mu_S^2 = \lambda_S\, v_S^2 + \lambda_{HS}\, v^2\,.\ee
Clearly, the large hierarchy between the $U(1)_X$ and electroweak breaking scales, i.e. $v_S \gg v$, can be arranged through $\mu_S^2 \gg \mu_H^2$ and $\lambda_{HS} \ll 1$. As it is well-known, such an arrangement is \emph{not technically natural} and would require fine-tuning in the absence of the implementation of any additional mechanism that resolves the gauge hierarchy problem. 

The neutral component of $H$ and real of $S$, to be identified as $\tilde{h}/\sqrt{2}$ and $\tilde{s}/\sqrt{2}$, respectively, lead to two physical neutral scalars,
\be \label{neut_scal}
h = \cos \phi\, \tilde{h} + \sin \phi\, \tilde{s}\,,\indent s = \cos \phi\, \tilde{s} - \sin \phi\, \tilde{h}\,,\ee
with
\be \label{phi}
\tan 2 \phi = \frac{2 \lambda_{HS}\, v\, v_S}{\lambda_H v^2 - \lambda_S v_S^2}\,.\ee
Apparently, the mixing is suppressed in the limit $v \ll v_S$ and $h$ is dominantly given by $\tilde{h}$. The state $h$ can be identified with the observed SM-like Higgs boson with mass $M_h=125.20 \pm 0.11$ GeV \cite{ATLAS:2012yve,CMS:2012qbp}. The other neutral scalar has a mass of the order of $U(1)_X$ breaking scale.

\subsection{Gauge sector}
\label{subsec:gauge}
This sector is also relatively simple, as the SM is extended by only a single abelian local symmetry. The covariant derivatives of $H$ and $S$ giving masses to the gauge bosons through their kinetic terms are,
\beqa \label{Co:derivative}
D_\mu H &=& \left(\partial_\mu + i g \frac{\tau^i}{2}  W^{i}_\mu + i \frac{ g'}{2} B_\mu + i g_X X^\prime_\mu\right)H\, , \nonumber \\
D_\mu S &=& \left( \partial_\mu - i  g_X X^\prime_\mu\right) S\,, \eeqa
since $H$ and $S$ are charged as $+1$ and $-1$ under the $U(1)_X$, respectively.

Substituting the VEVs given by Eq. (\ref{VEVs}), the neutral gauge boson mass matrix in the basis of $G^0=(B,W^3,X^\prime)^T$ is obtained as
\be \label{M_GB}
M_{G^0}^2 = \left(
	\ba{c c c}
	\frac{1}{4} g'^2 v^2 & -\frac{1}{4} g g' v^2 &  \frac{1}{2} g' g_X  v^2 \\
	- \frac{1}{4} g g'v^2& \frac{1}{4} g^2 v^2&  -\frac{1}{2} g  g_X  v^2\\
	\frac{1}{2} g' g_X  v^2&  -\frac{1}{2} g g_X  v^2& g_X^2 \left(v^2 + v_S^2 \right)\\
	\ea
	\right) \,.\ee
The above leads to one massless state, which is to be identified with the photon $A_\mu$, and two massive gauge bosons to be identified with physical $Z_\mu$ and $X_\mu$. Explicitly,
\be \label{GB_basis_change}
\left(\ba{c} B  \\  W^3 \\ X^\prime \ea \right)  = \left(\ba{ccc} \cos \theta_W & -\sin \theta_W & 0 \\
    \sin \theta_W & \cos \theta_W & 0\\
    0 & 0 & 1\\ \ea \right) \left(\ba{ccc} 1 & 0 & 0 \\
    0 & \cos \theta & - \sin \theta\\
    0 & \sin \theta & \cos \theta \\ \ea \right)
    \left( \ba{c} A  \\  Z\\ X \ea \right) \, ,\ee
with 
\be \label{GB_mixing}
\tan \theta_W = \frac{g^\prime}{g}\,,\indent \tan 2\theta = \frac{2\, m^2_{YX}}{m^2_{YY} - m^2_{XX}}\,,\ee
and
\be
m^2_{YY} = \frac{g^2}{\cos^2\theta_W} \frac{v^2}{4}\,,~~m^2_{XX}=g_X^2 \left(v_S^2 + v^2\right)\,,~~m^2_{YX} = - \frac{g g_X}{\cos\theta_W} \,\frac{v^2}{2}\,.\ee
The masses of $Z$ and $X$ bosons are obtained as
\be \label{M_XZ}
M^2_{Z, X} = \frac{1}{2} \left(m_{YY}^2 + m_{XX}^2 \mp  \sqrt{(m_{YY}^2 - m_{XX}^2)^2 +4\, m_{YX}^4} \right) \,.\ee

The gauge boson spectrum can be further simplified in the limit $v \ll v_S$. The $Z$-$X$ mixing angle becomes small, and it is given by
\be \label{th_apprx}
2\theta \approx \frac{g}{g_X \cos \theta_W}\frac{v^2}{v_S^2}\,.\ee
The neutral gauge boson masses can be approximated to
\be \label{ZX_apprx}
M_Z^2 \approx \frac{g^2 v^2}{4\cos^2\theta_W}\,,\indent M_X^2 \approx g_X^2 v_S^2\,\left(1+\frac{v^2}{v_S^2} \right)\,,\ee
with the corrections to the above arising only at ${\cal O}(v^4/v_S^4)$. The spectrum of the $W$-boson remains unchanged in the model.

\subsection{Yukawa sector}
\label{subsec:yukawa}
The mechanism discussed in the previous section can be implemented on the SM quarks and charged leptons through the following set of Yukawa interactions and mass terms between the chiral and chain fermions listed in Table \ref{tab:field}. 
\beqa \label{L:SM:Yukawa} 
-{\cal L}_Y &=& \sum_{\alpha=1}^{3} ((y_u)_\alpha\, \overline{q}_{L \alpha} \tilde{H} U_{R\alpha} + (y_u^\prime)_\alpha\, \overline{U}_{L (6-\alpha)} S^* u_{R \alpha}) + \sum_{i,j=1}^{5} \left(M^{(0)}_U\right)_{ij}\,\overline{U}_{Li} U_{Rj}\, \nonumber\\
&+& \sum_{\alpha=1}^{3} \left((y_d)_\alpha\, \overline{q}_{L \alpha} H D_{R\alpha} + (y_d^\prime)_\alpha\, \overline{D}_{L (6-\alpha)} S d_{R \alpha} \right) + \sum_{i,j=1}^{5} \left(M^{(0)}_D\right)_{ij}\,\overline{D}_{Li} D_{Rj}\, \nonumber\\
&+& \sum_{\alpha=1}^{3} \left((y_e)_\alpha\, \overline{l}_{L \alpha} H E_{R\alpha} + (y_e^\prime)_\alpha\,\overline{E}_{L (6-\alpha)} S e_{R \alpha} \right) + \sum_{i,j=1}^{5} \left( M^{(0)}_E\right)_{ij} \overline{E}_{Li} E_{Rj} \nonumber \\
&+& {\rm h.c.}\,, \eeqa
where, following Eq. (\ref{MFC}), the $5 \times 5$ tree-level mass matrices of the vectorlike fermions take the form
\be \label{MFC_F}
\left(M_F^{(0)}\right)_{ij} = W_F\,\left[\,\delta_{ij} + t_F\, (\delta_{i+1,j} + \delta_{i,j+1})\right]\,,\ee
with $F=U,D,E$. The presence of only electroweak doublet scalar enforces that $U_{Li,Ri}$ and $D_{Li,Ri}$ take opposite charges under the $U(1)_X$. It is also important to note that the direct Yukawa coupling between $q_{L3}$ and $u_{R3}$ or $d_{R3}$ is forbidden because of $U(1)_X$ even though they are allowed by the global symmetry group $G$ responsible for the latticised arrangement.

Without loss of generality, the parameters in $y_f$ and $y^\prime_f$, with $f = u,d,e$, can be made real by rephasing the right-chiral components of vectorlike and chiral fermions. $W_F$ can also be made real by rotating away their phases through left-chiral components of the chain fermions. This leaves only three physical phases arising through the off-site interaction parameters $t_F$. The number of free parameters in the Yukawa sector, therefore, is similar to the previous models of radiative mass mechanism \cite{Mohanta:2022seo,Mohanta:2024wcr} despite an enlarged vectorlike fermion sector.

The VEVs of the neutral components of $H$ and $S$ generate the mass-mixing between the chiral and chain fermions,
\be \label{mu_f}
\mu_{f \alpha} = (y_f)_\alpha\, \frac{v}{\sqrt{2}}\,, \indent  \mu^\prime_{f \alpha} = (y^\prime_f)_\alpha\, \frac{v_S}{\sqrt{2}}\,, \ee
leading to the structure hypothesised in Eq. (\ref{Lm}). Subsequently, when one-loop corrections are switched on, the full $8 \times 8$ fermion mass matrices take the form given in Eq. (\ref{calM}) for each sector. Namely,
\be \label{calMf}
{\cal M}_f = \left(\ba{cc} \left(0\right)_{3 \times 3} & (\mu_f)_{3\times5}
\\ (\mu_f^\prime)_{5\times3}
& \left(M_F\right)_{5 \times 5}\ea\right)\,,\ee
along with 
\be\label{MF:1loop} 
M_F = M_F^{(0)} + \delta M^{(0)}_F\,.\ee
Here, $\delta M^{(0)}_F$ is given by the simple and straightforward generalisation of Eq. (\ref{dM1}). ${\cal M}_f $ is diagonalised following the same procedure outlined in the previous section, with the following generalisation of notation:
\be \label{gen_U}
\rho_L \to \rho_{f L} = -\mu_f M_F^{-1},~\rho^\dagger_R \to (\rho_{fR})^\dagger = -M_F^{-1} \mu_f^\prime, ~u_{L,R} \to u_{fL,fR},~U_{L,R} \to U_{fL,fR}.\ee

Upon integrating out the heavy fermions, the effective $3 \times 3$ charged fermion mass matrices at the leading order become 
\be \label{mf_eff} 
m^f_{\rm eff} \simeq -\mu_f\, M_F^{-1}\, \mu_f^{\prime} \equiv M_f\,, \ee
with subleading corrections of ${\cal O}(\rho_{L,R}^3)$. The diagonalisation of $M_f$, through
\be \label{mf_diag}
u_{fL}^\dagger\, M_f\, u_{fR} = {\rm Diag}(m_{f1},m_{f2},m_{f3})\,,\ee
result in the masses of light fermions.

\subsection{Asymptotic freedom and Landau poles}
\label{subset:AF}
Among the new fields introduced, the vectorlike fermions modify the running of the gauge couplings. At 1-loop, the renormalisation group equations of the gauge couplings, $g_i \equiv \sqrt{4 \pi \alpha_i}$, are given by
\be \label{gi_RGE}
\frac{d \alpha_i}{d \ln \mu} = \frac{\alpha_i^2}{2 \pi}\, b_i\,.\ee
The beta-function coefficient $b_i$ for the SM gauge couplings in the presence of $N$ copies of chain fermions listed in Table \ref{tab:field} are obtained as \cite{Machacek:1983tz},
\be \label{bi}
b_3=-7+\frac{4}{3}N\,,\indent b_2 = -\frac{19}{6}\,,\indent b_1=\frac{41}{10}+\frac{32}{15} N\,.\ee
The $SU(2)_L$ beta function does not get modified as we introduced only the weak isosinglet fermions. The running of $U(1)_X$ gauge coupling is given by
\be \label{bx}
b_X = \frac{2}{3} + \frac{28}{3} N + \frac{1}{3}\,,\ee
where the first term is a contribution from the Higgs doublet, while the second and third arise from the chain fermions and $S$, respectively.

It is noteworthy that $N=5$ is the maximum value for which $b_3$ remains negative. Coincidentally, this value matches the minimum number of chain-fermions one requires to produce a rank-one mass matrix for the light fermions. Consequently, the QCD at 1-loop remains asymptotically free in the present framework.

The large contribution to $b_{1,X}$ by the new fermions indicates that the $U(1)$ Landau poles are reached at much lower scales. Typically, if the chain-fermion mass scale is of the order of $m_F$, then
\be
\Lambda_{\rm Landau} \simeq m_F\,\exp\left(\frac{2 \pi}{b_{1,X}\, \alpha_{1,X}[m_F]} \right)\,.\ee
For $N=5$, $m_F = 10~\mathrm{TeV}$, and assuming $\alpha_X = \alpha_1$ at the scale $m_F$, we find that the Landau poles appear at $\Lambda_{\rm Landau} \sim 10^{4}\,\mathrm{TeV}$ for $U(1)_X$ and at $\Lambda_{\rm Landau} \sim 10^{11}\,\mathrm{TeV}$ for $U(1)_Y$. Evidently, the $U(1)_X$ gauge coupling becomes non-perturbative at a scale only two to three orders of magnitude above $m_F$. This indicates that, while the model provides a consistent, effective description at energies $m_F$ and below, it cannot be reliably extrapolated to significantly higher scales. The theory, and in particular the new sector, becomes strongly coupled shortly above $m_F$. While this does not affect the physics below the scale $m_F$, this limitation could be avoided by embedding the $U(1)_X$ gauge symmetry into a non-Abelian gauge theory.

The running of the Higgs quartic couplings and a possibility to turn negative at a scale above $m_F$ can also put a constraint on the cut-off scale of the theory. In the present model, neither $H$ nor $S$ couples to the pair of new fermions through direct coupling due to the underlying gauge invariance. In this case, the 1-loop beta function coefficients of the quartic couplings do not get direct contributions from the chain-fermions. The latter, however, affect the running of $\lambda_{H}$ through modification in the running behaviour of the top Yukawa coupling and $g_3$. Since $g_3$ is larger at large $\mu_R$ in comparison to that in the SM, it increases the rate at which the top-quark Yukawa coupling decreases. Subsequently, the latter provides somewhat lesser negative contribution to the beta function of the quartic coupling, allowing it to remain positive at larger $\mu_R$. As it is shown in \cite{Gopalakrishna:2018uxn}, this effect can even change the direction of running of the quartic coupling for $N=5$, allowing it to stay positive. Nevertheless, the appearance of the Landau pole shortly above $m_F$ implies that the theory should be regarded as an effective description, and the stability of the vacuum within this limited range of scales is preserved.

\section{BSM Effects}
\label{sec:bsm}
The minimal embedding of the latticed theory space construction in the SM implies the existence of an extended sector with five copies of vectorlike fermions for each of the three sectors, a vector and a scalar boson. They lead to novel phenomenological effects both through their direct couplings with the SM fermions and by modifying the existing SM couplings. We derive the most relevant set of these below to assess the phenomenological implications in the subsequent sections.

\subsection{Couplings with $X$-boson}
\label{subsec:BSMX}
The analogue of Eq. (\ref{L:gauge}) along with the mixing effects through Eq. (\ref{GB_basis_change}) lead to the following interaction between the physical $X$-boson and heavy and light fermions in the model,
\be \label{LX0}
{\cal L}_X = g_X c_\theta\,J_X^\mu X_\mu - \frac{g s_\theta}{c_W}\,  \left(J_{Zf}^\mu + J_{ZF}^\mu\right) X_\mu\,. \ee
Here and henceforth, we use $c_\theta = \cos \theta$, $s_\theta = \sin \theta$, $c_W = \cos \theta_W$ and so on. The associated currents are given by,
\beqa \label{Js}
J^\mu_{Zf} &=& \sum_{f,\alpha} \left(\left(T^f_3 - s^2_W Q_f\right) \,\bar{f}_{L \alpha} \gamma^\mu f_{L \alpha} - s_W^2 Q_f\, \bar{f}_{R \alpha} \gamma^\mu f_{R \alpha} \right)\,, \nonumber \\
J^\mu_{ZF} &=& - \sum_{F,i} s^2_W Q_F\,\left(\bar{F}_{L i} \gamma^\mu F_{L i} + \bar{F}_{R i} \gamma^\mu F_{R i} \right)\,, \nonumber \\
J^\mu_{X} &=& \sum_{F,i} q_F\,\left(\bar{F}_{L i} \gamma^\mu F_{L i} + \bar{F}_{R i} \gamma^\mu F_{R i} \right)\,.\eeqa
Here, $Q_{f,F}$ are electromagnetic charges of the light and heavy fermions of type $f$ or $F$, respectively, while $q_F$ is the charge under the $U(1)_X$, see Table \ref{tab:field}. 

To obtain the effective couplings with the physical SM fermions, we perform the basis change following Eq. (\ref{88U_block}) and find,
\be \label{LX1}
{\cal L}_X \supset  \sum_{f,\alpha,\beta} \left( \left(g_{f_L}^X \right)_{\alpha \beta}\, \bar{f}_{L \alpha} \gamma^\mu f_{L \beta} + \left(g_{f_R}^X \right)_{\alpha \beta}\, \bar{f}_{R \alpha} \gamma^\mu f_{R \beta} \right)  X_\mu\,,\ee
where we now denote the physical fermion fields without primed notation. The corresponding couplings are determined as,
\beqa \label{dgX}
\left(g_{f_L}^X \right)_{\alpha \beta} &=& g_X q_F\, c_\theta \left(u_{fL}^\dagger \rho_{fL} \rho_{fL}^\dagger u_{fL} \right)_{\alpha \beta} - \frac{g s_\theta}{c_W} \left(T_3^f-s_W^2 Q_f \right) \delta_{\alpha \beta}\,, \nonumber \\
\left(g_{f_R}^X \right)_{\alpha \beta} &=& g_X q_F\, c_\theta \left(u_{fR}^\dagger \rho_{fR} \rho_{fR}^\dagger u_{fR} \right)_{\alpha \beta} + \frac{g s_\theta}{c_W} s_W^2 Q_f\, \delta_{\alpha \beta}\,.\eeqa
Consequently, the $X$-boson has flavour-violating couplings with the SM fermions which arise at ${\cal O}(\rho_{L,R}^2)$ through heavy-light fermion mixings. The $Z$-$X$ mixing also contributes to the flavour universal couplings; however, they are suppressed by a factor of ${\cal O}(v^2/v_S^2)$ as it can be deduced from the above and Eq. (\ref{th_apprx}).

\subsection{Modification to $Z$-boson couplings}
\label{subsec:BSMZ}
The interaction Lagrangian involving the $Z$-boson modifies to,
\be \label{LZ0}
{\cal L}_Z = \frac{g c_\theta}{c_W}\, \left(J_{Zf}^\mu + J_{ZF}^\mu\right) Z_\mu\, +\, g_X s_\theta\,J_X^\mu Z_\mu\,. \ee
While the first term, as usual, leads to flavour-conserving couplings of the SM fermions with the $Z$-boson, the second term gives rise to flavour-changing couplings through heavy-light mixing. An additional contribution also arises through the third term, which is doubly suppressed by small $Z$-$X$ and heavy-light fermion mixing.

Converting into the physical basis of fermions and neglecting the doubly suppressed contributions, we find 
\be \label{LZ1}
{\cal L}_Z \supset \frac{g c_\theta}{c_W}\,  \left(J_{Zf}^\mu  +  \sum_{f,\alpha,\beta} \left( \left(\delta g_{f_L}^Z \right)_{\alpha \beta}\, \bar{f}_{L \alpha} \gamma^\mu f_{L \beta} + \left(\delta g_{f_R}^Z \right)_{\alpha \beta}\, \bar{f}_{R \alpha} \gamma^\mu f_{R \beta} \right) \right)  Z_\mu\,,\ee
for the $Z$-boson interaction with the physical SM fields, where
\be \label{dgZ}
\left(\delta g_{f_L}^Z \right)_{\alpha \beta} = - T_3^f \left(u^\dagger_{fL} \rho_{fL} \rho^\dagger_{fL} u_{fL} \right)_{\alpha \beta}\,,\indent \left(\delta g_{f_R}^Z \right)_{\alpha \beta} = 0\,,\ee 
denote the deviation from the standard couplings with the $Z$-boson. It is noticed that only the couplings with left-handed fields get modified. This is a consequence of adding only the $SU(2)_L$ singlet vectorlike fermions.

\subsection{Modification to Higgs couplings}
\label{subsec:BSMH}
Unlike in the SM, the $SU(2)_L$ doublet Higgs does not couple to the right-chiral SM fields directly in the present model. Their couplings to the SM fields, therefore, arise effectively when heavy fermions are integrated out. We begin with the interaction representative of Eq. (\ref{L:SM:Yukawa}) which after the $U(1)_X$ and electroweak symmetry breaking takes the form:
\be \label{LY0}
-{\cal L}_Y \supset \sum_{\alpha} \frac{1}{\sqrt{2}} \left((y_f)_\alpha\, \bar{f}_{L \alpha} F_{R \alpha} \, \tilde{h}\, + (y^\prime_f)_\alpha\,\overline{F}_{L(6-\alpha)} f_{R \alpha}\,\tilde{s} \right) + {\rm h.c.}\,. \ee
Here, $\tilde{h}$ and $\tilde{s}$ are to be replaced with their physical counterparts, following Eq. (\ref{neut_scal}), to obtain the fermion couplings with the latter.

Next, we integrate out the heavy fermions using Eq. (\ref{88U_block}) and compute the effective interaction in the physical basis of fermions and the Higgs boson. It is obtained as,
\be \label{LY1}
-{\cal L}_Y \supset  \sum_{\alpha} \frac{m_{f \alpha}}{v}\,\left(c_\phi\,+ \frac{v}{v_S}\,s_\phi\right)\,\bar{f}_{L \alpha} f_{R \alpha}\,h\,\,+\,\sum_{\alpha,\beta} \left(\delta g_f^h \right)_{\alpha \beta}\,\bar{f}_{L \alpha} f_{R \beta}\,h\, + \, {\rm h.c.}\,,\ee
with
\be \label{dgh}
\left(\delta g_f^h \right)_{\alpha \beta} = -\frac{1}{2} \left(\frac{c_\phi\,m_{f\beta} }{v} \left( u_{fL}^\dagger \rho_{fL} \rho_{fL}^\dagger u_{fL} \right)_{\alpha \beta} + \frac{s_\phi\, m_{f\alpha}  }{v_S} \left( u_{fR}^\dagger \rho_{fR} \rho_{fR}^\dagger u_{fR} \right)_{\alpha \beta}\right)\,,\ee
As a result, the Higgs boson's couplings with the SM fermions are identical to those in the SM at the leading order, and the deviation arises only at ${\cal O}(\rho_{L,R}^2)$. Also, the contribution to the deviation emerging from the right-handed fermion mixing is further suppressed by a small $h$-$s$ coupling, see Eq.(\ref{phi}).

\subsection{FCNC coupling structure}
\label{subsec:BSMFCNC}
It is observed from Eqs. (\ref{dgX},\ref{dgZ},\ref{dgh}) that all the flavour-changing effects are proportional to the off-diagonal elements of the combinations,
\be \label{omega}
u_{fL}^\dagger \rho_{fL} \rho_{fL}^\dagger u_{fL} \equiv \Omega_{fL}\,~~~{\rm or}~~~~ u_{fR}^\dagger \rho_{fR} \rho_{fR}^\dagger u_{fR}\equiv \Omega_{fR}\,.\ee
Therefore, it is a noteworthy exercise to investigate the flavour pattern of these couplings. These matrices can be written as sums of two qualitatively distinct contributions using the straightforward decomposition, 
\beqa \label{FVL_decomp}
\left( \Omega_{fL} \right)_{\alpha \beta} = \left( u_{fL}^\dagger \rho_{fL} \rho_{fL}^\dagger u_{fL} \right)_{\alpha \beta} &=& \sum_{m=1}^2 \left(u_{fL}^\dagger \rho_{fL}\right)_{\alpha m} \left(u_{fL}^\dagger \rho_{fL}\right)^*_{\beta m}\nonumber \\ 
&+& \sum_{\gamma=1}^3 \left(u_{fL}^\dagger \rho_{fL}\right)_{\alpha, (6-\gamma)} \left(u_{fL}^\dagger \rho_{fL}\right)^*_{\beta,(6-\gamma)}\,,
\eeqa
and
\beqa \label{FVR_decomp}
\left( \Omega_{fR} \right)_{\alpha \beta} = \left( u_{fR}^\dagger \rho_{fR} \rho_{fR}^\dagger u_{fR} \right)_{\alpha \beta} &=& \sum_{m=4}^5 \left(u_{fR}^\dagger \rho_{fR}\right)_{\alpha m} \left(u_{fR}^\dagger \rho_{fR}\right)^*_{\beta m}\nonumber \\ 
&+& \sum_{\gamma=1}^3 \left(u_{fR}^\dagger \rho_{fR}\right)_{\alpha \gamma} \left(u_{fR}^\dagger \rho_{fR}\right)^*_{\beta \gamma}\,.
\eeqa
We show below that the second terms in both these equations are directly linked to the SM flavour pattern, while the contributions captured by the first terms depend primarily on the off-site coupling strength in the lattice theory space.

Using the definition of the light fermion mass matrix $M_f$ given in Eq. (\ref{mf_eff}) and Eq. (\ref{gen_U}), it can be shown that
\beqa \label{MMdag}
\left(u_{fL}^\dagger M_f M_f^\dagger u_{fL} \right)_{\alpha \beta} &=&  \sum_{\gamma=1}^3\,|\mu_{f\gamma}^\prime|^2 \, \left(u_{fL}^\dagger \rho_{fL}\right)_{\alpha, (6-\gamma)} \left(u_{fL}^\dagger \rho_{fL}\right)^*_{\beta,(6-\gamma)}\,.
\eeqa
The above, through definition Eq. (\ref{mf_diag}), must be proportional to a diagonal matrix, non-vanishing elements of which are the SM fermion masses. For generic and non-hierarchical values of $|\mu_{f \gamma}^\prime|$, the SM flavour pattern must be encoded in the relevant elements of $u_{fL}^\dagger \rho_{fL}$. Therefore, one typically expects
\be \label{gim_L}
\left(u_{fL}^\dagger \rho_{fL}\right)_{\alpha, (6-\gamma)}  \sim\, {\cal O}\left(\frac{m_{f \alpha}}{|\mu_f^\prime|}\right)\, \quad {\rm for}\,\gamma=1,2,3\,.
\ee
Similarly, an identical expectation from the diagonalisation of $M_f^\dagger M_f$ leads to,
\be \label{gim_R}
\left(u_{fR}^\dagger \rho_{fR}\right)_{\alpha \gamma}  \sim\, {\cal O}\left(\frac{m_{f \alpha}}{|\mu_f|}\right)\, \quad {\rm for}\,\gamma=1,2,3\,.
\ee
The above, when substituted into the second terms of Eqs. (\ref{FVL_decomp},\ref{FVR_decomp}), implies that the corresponding flavour-changing couplings are proportional to ${\cal O}\left(m_\alpha m_\beta/|\mu_f^\prime|^2\right)$ and ${\cal O}\left(m_\alpha m_\beta/|\mu_f|^2\right)$, respectively. This is equivalent to a GIM-like \cite{Glashow:1970gm} suppression, which has also been shown to generically exist in the models of seesaw-induced flavour hierarchies \cite{Joshipura:1989xa}.

The contributions to FCNC couplings arising from the first terms in Eqs. (\ref{FVL_decomp},\ref{FVR_decomp}) does not possess GIM-like suppression in general. Their size and structure can be estimated in the leading order approximation, in which,
\beqa \label{non-gim}
\left(u_{fL}^\dagger \rho_{fL}\right)_{\alpha m} & \approx & - \sum_{\beta=1}^3 \left(u_{fL}^\dagger \right)_{\alpha \beta}\,\mu_{f \beta}\,\left(M_F^{(0)}\right)^{-1}_{\beta m}\,, \nonumber \\
\left(u_{fR}^\dagger \rho_{fR}\right)^*_{\alpha m} & \approx & - \sum_{\beta=1}^3 \left(u_{fR}^\dagger\right)_{\alpha \beta} \mu^\prime_{f\beta}\,\left(M_F^{(0)}\right)^{-1}_{m,6-\beta}\,,\eeqa
and 
\beqa \label{M_inv}
\left(M_F^{(0)}\right)^{-1} &=& \frac{1}{W_F\, (1-4 t_F^2 + 3 t_F^4)} \times \nonumber \\
& &\left(
\begin{array}{ccccc}
1 - 3 t_F^2 + t_F^4
& -t_F + 2 t_F^3
& t_F^2 - t_F^4
& -t_F^3
& t_F^4 \\[6pt]
 -t_F + 2 t_F^3
& 1 - 2 t_F^2
& -t_F + t_F^3
& t_F^2
& -t_F^3 \\[6pt]
 t_F^2 - t_F^4
& -t_F + t_F^3
& 1 - 2 t_F^2 + t_F^4
& -t_F + t_F^3
& t_F^2 - t_F^4 \\[6pt]
 -t_F^3
& t_F^2
& -t_F + t_F^3
& 1 - 2 t_F^2
& -t_F + 2 t_F^3 \\[6pt]
 t_F^4
& -t_F^3
& t_F^2 - t_F^4
& -t_F + 2 t_F^3
& 1 - 3 t_F^2 + t_F^4
\end{array}
\right). \eeqa
Apparently, the off-diagonal elements of $(M_F^{(0)})^{-1}$ decrease progressively when $|t_F| < 1$. Next-to-leading-order effects introduce only small corrections to $(M_F^{(0)})^{-1}$, while leaving its hierarchical structure unaltered.

Combining both the contributions, the flavour structure of FCNC couplings can be characterised as,
\beqa \label{FV_decomp}
\left(\Omega_{fL}\right)_{\alpha \beta}  &\sim & {\cal O} \left(\frac{|\mu_f|^2}{W_F^2}\right) \left[u_{fL}^\dagger \left(\ba{ccc} {\cal O}(1) & {\cal O}(t_F) & {\cal O}(t_F^2) \\
{\cal O}(t_F) & {\cal O}(1) & {\cal O}(t_F) \\
{\cal O}(t_F^2) & {\cal O}(t_F) & {\cal O}(t_F^2) \ea \right) u_{fL}\right]_{\alpha \beta} + {\cal O}\left(\frac{m_\alpha m_\beta}{|\mu_f^\prime|^2}\right)\,, \nonumber \\
\left(\Omega_{fR}\right)_{\alpha \beta} &\sim & {\cal O} \left(\frac{|\mu^\prime_f|^2}{W_F^2}\right) \left[u_{fR}^\dagger \left(\ba{ccc} {\cal O}(1) & {\cal O}(t_F) & {\cal O}(t_F^2) \\
{\cal O}(t_F) & {\cal O}(1) & {\cal O}(t_F) \\
{\cal O}(t_F^2) & {\cal O}(t_F) & {\cal O}(t_F^2) \ea \right) u_{fR}\right]_{\alpha \beta} + {\cal O}\left(\frac{m_\alpha m_\beta}{|\mu_f|^2}\right)\,.
\eeqa
Certain aspects of the potential FCNC effects can already be qualitatively understood from the above results. For example, the FCNC induced in the right-handed sector is less suppressed in comparison to that in the left-handed sector. They primarily originate from non-GIM-like contributions. Moreover, these effects are expected to be the largest in the up-type quark sector since the top-quark mass is very close to the electroweak scale and inducing it through seesaw would require $\mu_u^\prime \sim W_U$ as well as $y_{ui} \sim {\cal O}(1)$.

\section{Benchmark Solutions}
\label{sec:benchmark}
We now examine the viability of the framework in reproducing the observed pattern of charged fermion masses and quark mixing parameters. The usual $\chi^2$-optimisation method is used to carry this out, where
\be \label{chs}
\chi^2 = \sum_i \left(\frac{O^{\rm th}_i - O^{\rm exp}_i}{\sigma^{\rm exp}_i} \right)^2\,.\ee
Here, $O^{\rm exp}_i$ are the values of the charged fermion masses and quark mixing parameters extrapolated to a renormalisation scale $\mu_R$. $\sigma^{\rm exp}_i$ are the corresponding $1\sigma$ uncertainties evolved to the same scale. For reference, we use the values of these observables at $\mu_R = 10$ TeV derived from \cite{Antusch:2025fpm}, which uses two-loop renormalisation group equations (RGE) to evolve them to a given $\mu_R$ from their corresponding low-energy values from PDG 2024 \cite{ParticleDataGroup:2024cfk}. These values are reproduced in Table \ref{tab:input}.
\begin{table}[t]
\centering
\begin{tabular}{ccc}
\toprule
Sector & Observable & Value \\
\midrule
\multirow{3}{*}{Up-type quarks}
& $m_t$ & $(137.2 \pm 0.8)\,\text{GeV}$ \\
& $m_c$ & $(487 \pm 14)\,\text{MeV}$ \\
& $m_u$ & $(0.97 \pm 0.21)\,\text{MeV}$ \\
\addlinespace
\hline
\addlinespace
\multirow{3}{*}{Down-type quarks}
& $m_b$ & $(2.171 \pm 0.021)\,\text{GeV}$ \\
& $m_s$ & $(42.8 \pm 3.5)\,\text{MeV}$ \\
& $m_d$ & $(2.14 \pm 0.19)\,\text{MeV}$ \\
\addlinespace
\hline
\addlinespace
\multirow{3}{*}{Charged leptons}
& $m_\tau$ & $(1.773 \pm 0.010)\,\text{GeV}$ \\
& $m_\mu$ & $(104.4 \pm 0.1)\,\text{MeV}$ \\
& $m_e$ & $(0.4957 \pm 0.0003)\,\text{MeV}$ \\
\addlinespace
\hline
\addlinespace
\multirow{4}{*}{Quark mixing}
& $V^{\rm CKM}_{12}$ & $0.2251 \pm 0.0008$ \\
& $V^{\rm CKM}_{23}$ & $0.043 \pm 0.004$ \\
& $V^{\rm CKM}_{13}$ & $0.0038 \pm 0.0008$ \\
& $J_{\rm CP}$ & $(3.33 \pm 0.09)\times 10^{-5}$ \\
\bottomrule
\end{tabular}
\caption{The reference values of the quark and charged lepton masses and CKM parameters at the renormalisation scale $\mu_R=10$ TeV used in numerical fit.}
\label{tab:input}
\end{table}

$O^{\rm th}_i$ are the theoretical values of the observables computed within the framework as functions of the input parameters. They are numerically evaluated using $M_f$ defined in Eq. (\ref{mf_eff}) for $f=u,d,e$. The diagonalisation of $M_f$ yields the physical fermion masses, while the Cabibbo–Kobayashi–Maskawa (CKM) matrix is computed as $V^{\rm CKM} = u_{uL}^\dagger u_{dL}$ from the unitary matrices obtained from Eq. (\ref{mf_diag}). As outlined in Section \ref{subsec:yukawa}, the parameters $\mu_{f \alpha}$, $\mu^\prime_{f \alpha}$, and $W_F$ can be chosen real by suitable field redefinitions of the SM and chain fermion fields, while $t_F$ remains complex in general. The flavour spectrum also depends on $g_X$ and $M_X$ through next-to-leading-order contributions to $M_f$. In this way, the thirteen SM observables are determined by a total of 29 real parameters in the present framework.

Although the number of parameters is much larger than the number of observables, it is not obvious that a good fit can always be found. This is primarily because the observables are complex non-linear functions of the parameters, and the dependence is not straightforward. Secondly, the dimensionless parameters are expected to take non-hierarchical ${\cal O}(1)$ values, unlike the SM Yukawa couplings, which are forced to be hierarchical. Moreover, the actual number of observable quantities exceeds those in the SM once the masses and couplings of vectorlike fermions are included; the latter are also largely determined by the fit, as we show below.

We fix $g_X = 1$ and numerically minimise $\chi^2$ for different values of $M_X$, while optimising the remaining parameters. During the minimisation, we require that the dimensionless couplings satisfy a perturbativity limit, $|y_{f\alpha}|$, $|y^\prime_{f\alpha}|$, $|t_F| < \sqrt{8\pi/3}$. The latter has been obtained by considering contributions to tree-level $2 \to 2$ scatterings in the high-energy limit \cite{Allwicher:2021rtd}, and it is more stringent than the typical bound $y < 4\pi$ derived from the requirement that loop corrections remain subleading. For the dimensionful parameters $W_F$ representing overall scales of vectorlike fermions, we restrict their values in the vicinity of $M_X$, although they are not rigidly fixed during the minimisation. The VEV $v_S$ is determined from $g_X$ and $M_X$ from Eq. (\ref{ZX_apprx}). 

We find that viable solutions can be obtained for almost any value of $M_X \geq 1$ TeV for which all the aforementioned 13 observables can be reproduced within their $3\sigma$ ranges. This can be simply understood from the fact that the SM fermion masses scale as, $\mu_f^\prime/W_F$ and hence one can set arbitrarily large $W_F \sim M_X$ and $\mu_f^\prime$ but keeping the ratio fixed. We select three representative benchmark solutions corresponding to $M_X = 1$, $5$, and $10~\mathrm{TeV}$ to illustrate their characteristic features and associated phenomenological implications. The optimised parameter sets for these benchmarks, labelled as S1, S2, and S3, are given in Table~\ref{tab:fit_results}. The solution S1 reproduces all observables within the $1\sigma$ ranges listed in Table~\ref{tab:input}, with the sole exception of $m_b$, which deviates by only $1.2\sigma$. In contrast, the S2 and S3 solutions yield $\chi^2_{\rm min} \ll 1$ and reproduce all observables essentially to their central values.
\begin{table}[t]
\centering
\begin{tabular}{cccc}
\toprule
Parameter & S1 & S2 & S3 \\
\midrule
$M_X$          
& $1\times 10^{3}$ 
& $5\times 10^{3}$ 
& $1\times 10^{4}$ \\

\addlinespace
$W_U$          
& $2.2275\times 10^{3}$ 
& $7.4356\times 10^{3}$ 
& $2.0468\times 10^{4}$ \\
$W_D$          
& $1.3785\times 10^{4}$ 
& $5.5603\times 10^{4}$ 
& $3.7524\times 10^{5}$ \\
$W_E$          
& $1.2654\times 10^{4}$ 
& $4.5599\times 10^{4}$ 
& $8.5562\times 10^{4}$ \\

\addlinespace
$t_U$          
& $-0.97406 + 0.22612\,i$ 
& $0.92063 - 0.32115\,i$ 
& $0.90288 + 0.28292\,i$ \\
$t_D$          
& $0.96155 - 0.25501\,i$ 
& $-0.94969 + 0.31311\,i$ 
& $0.92170 + 0.26315\,i$ \\
$t_E$          
& $-0.98464 + 0.056430\,i$ 
& $0.98916 + 0.049848\,i$ 
& $0.97364 - 0.056887\,i$ \\

\midrule
$\mu_{u1}$     
& $160.91$ 
& $-154.38$ 
& $175.67$ \\
$\mu_{u2}$     
& $-200.42$ 
& $214.71$ 
& $266.54$ \\
$\mu_{u3}$     
& $-468.23$ 
& $-366.25$ 
& $-431.69$ \\

\addlinespace
$\mu'_{u1}$    
& $-704.34$ 
& $2.8361\times 10^{3}$ 
& $-6.5243\times 10^{3}$ \\
$\mu'_{u2}$    
& $-705.15$ 
& $-3.3540\times 10^{3}$ 
& $5.9816\times 10^{3}$ \\
$\mu'_{u3}$    
& $1.9544$ 
& $6.9431$ 
& $-13.052$ \\

\midrule
$\mu_{d1}$     
& $-45.449$ 
& $-68.678$ 
& $-57.186$ \\
$\mu_{d2}$     
& $-55.112$ 
& $-93.576$ 
& $-86.172$ \\
$\mu_{d3}$     
& $123.67$ 
& $-167.36$ 
& $139.02$ \\

\addlinespace
$\mu'_{d1}$    
& $213.37$ 
& $-401.06$ 
& $-5.2015\times 10^{3}$ \\
$\mu'_{d2}$    
& $191.58$ 
& $-687.41$ 
& $2.6426\times 10^{3}$ \\
$\mu'_{d3}$    
& $439.24$ 
& $-1.0884\times 10^{3}$ 
& $5.4504\times 10^{3}$ \\

\midrule
$\mu_{e1}$     
& $-22.151$ 
& $13.327$ 
& $-26.028$ \\
$\mu_{e2}$     
& $-22.016$ 
& $-12.790$ 
& $25.225$ \\
$\mu_{e3}$     
& $-169.67$ 
& $-114.24$ 
& $-169.25$ \\

\addlinespace
$\mu'_{e1}$    
& $-94.749$ 
& $-506.69$ 
& $505.75$ \\
$\mu'_{e2}$    
& $-91.206$ 
& $-436.48$ 
& $-527.15$ \\
$\mu'_{e3}$    
& $706.20$ 
& $-3.4530\times 10^{3}$ 
& $3.6499\times 10^{3}$ \\
\bottomrule
\end{tabular}
\caption{Optimised values of various model parameters for three benchmark solutions corresponding to $M_X=1$, $5$ and $10$ TeV. The values of dimensionful parameters are given in GeV.}
\label{tab:fit_results}
\end{table}

The most notable feature among the solutions in Table \ref{tab:fit_results} is that they all correspond to $W_U \ll W_D, W_E$. This arises from the fact that the third-generation masses are generated through a tree-level seesaw mechanism, and therefore the hierarchy $m_t \gg m_b, m_\tau$ translates into an inverted ordering among the mass scales of their vectorlike partners. One therefore typically expects the top partners to be the lightest new-physics states in this framework. The large top-quark mass and its seesaw origin also keep all the $\mu_{ui}$ very close to the electroweak scale. The magnitudes of $\mu_{di}$ and $\mu_{ei}$ are relatively smaller; however, they do not exhibit a strong intergenerational hierarchy, indicating that all the Yukawa couplings satisfy $|y_{fi}| \sim \mathcal{O}(1)$. An almost identical pattern is found for $y_{fi}^\prime$ as well, except in the up-type quark sector, for which $|y_{u3}|$ is found to be small. Nevertheless, this hierarchy is still significantly smaller in comparison to the five to six orders of magnitude range required in the SM without any additional mechanism. All the solutions also lead to $|t_F| \approx 1$, indicating the strong preference for universal nearest-neighbour off-site interaction strength in the latticed space.  

For all three solutions, we compute the magnitudes of $\Omega_{fL}$, $\Omega_{fR}$ matrices as defined in Eqs. (\ref{omega}). As mentioned in the previous section, they quantify BSM effects in the couplings of Higgs and vector bosons. The absolute values are listed in Appendix \ref{app:omega}. As anticipated already in Section \ref{subsec:BSMFCNC}, the largest couplings among these are found in $\Omega_{uR}$ indicating the dominant unitarity violation in the up-type quark sector. They also show a hierarchical structure with $(3,3)$ being the largest element and others decreasing as one goes away from the third-generation. The remaining of $\Omega_{fR}$ and $\Omega_{fL}$ are comparatively small and anarchic. For all the solutions $|\Omega_{fL}| \ll |\Omega_{fR}|$ as well as $|\Omega_{dL}|, |\Omega_{eL}| \ll |\Omega_{uL}|$ and $|\Omega_{dR}|, |\Omega_{eR}| \ll |\Omega_{uR}|$. All these features are along the expected lines as inferred from Eq. (\ref{FV_decomp}).

\section{Phenomenological Constraints}
\label{sec:pheno}
While reproducing the SM flavour spectrum does not strongly constrain the mass scales of the new gauge boson and the vectorlike fermions, these scales are nonetheless subject to lower bounds. The most stringent constraints are expected to arise from current direct search limits, since FCNC effects are highly suppressed in the present framework. This suppression is particularly pronounced in the down-quark and charged-lepton sectors as discussed in the section \ref{sec:bsm}. In contrast, FCNC processes in the up-quark sector may still provide meaningful constraints on the scale of new physics. In the present section, we discuss the most relevant constraints from the direct and indirect searches in the context of the three sample solutions. For explicit numerical computation, we set $c_\theta = c_\phi =1$ wherever their values are required.

\subsection{Direct searches}
\label{subsec:direct_search}
\begin{figure}[t]
\centering
\includegraphics[width=0.33\textwidth]{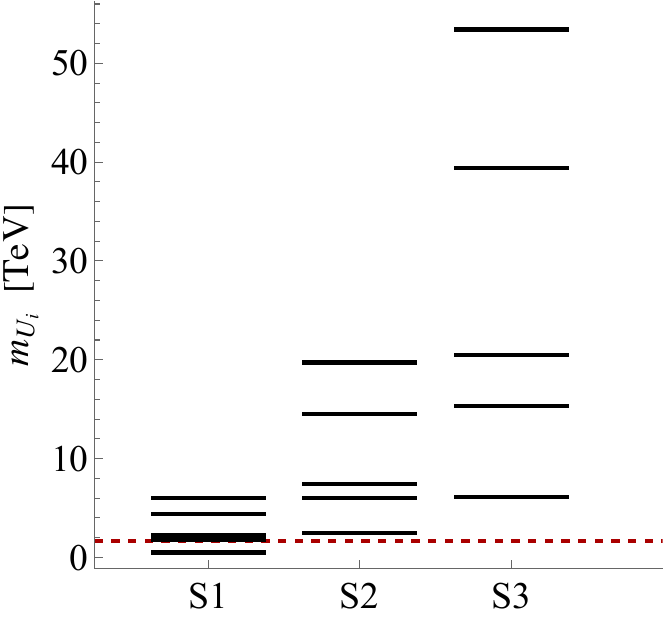}\hfill
\includegraphics[width=0.33\textwidth]{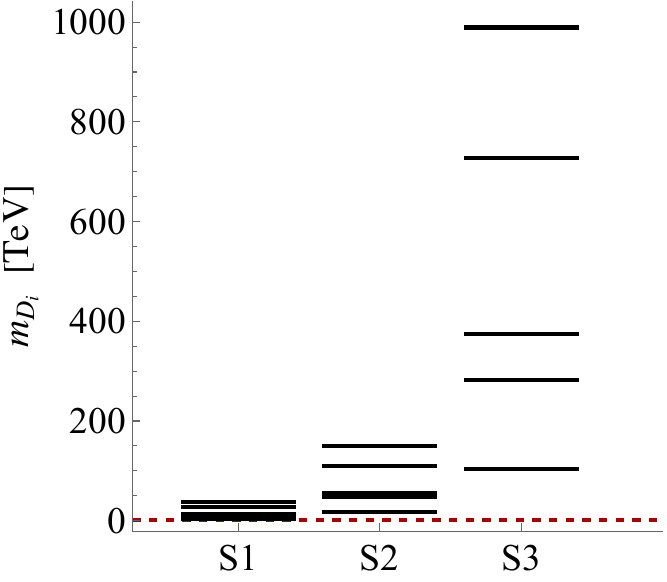}\hfill
\includegraphics[width=0.33\textwidth]{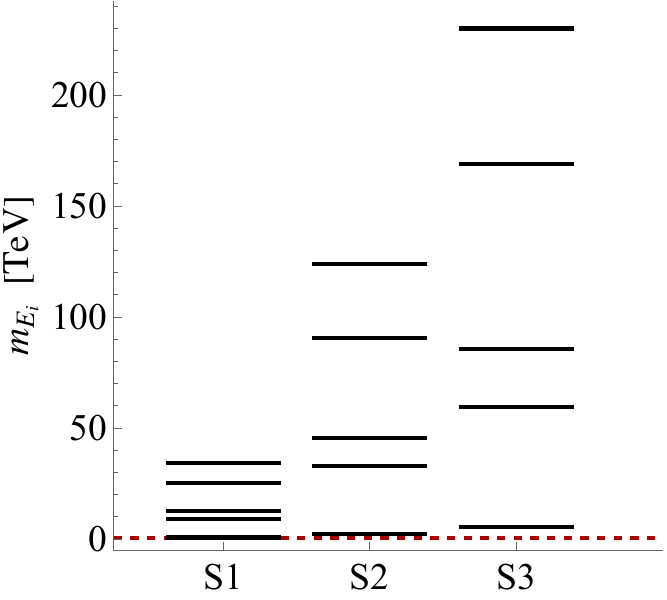}
\caption{Mass spectra of the vector-like fermions $U_i$, $D_i$, and $E_i$ for the benchmark solutions. The red dashed lines indicate the current most stringent direct-search exclusion limits for each type of vectorlike fermion.}
\label{fig4}
\end{figure}

The chain fermion mass spectrum computed for the three sample solutions is shown in Fig.~\ref{fig4}. In each case, the spectrum exhibits a reasonably wide spread around the scale $W_F$. This can be attributed to the fact that $|t_F| \approx 1$ in all solutions, which leads to significant deviations from a degenerate spectrum via Eq.~(\ref{m0_k}). Some of the vectorlike states remain well below $M_X$, particularly in the up-type quark and charged-lepton sectors, as seen in the spectrum. These states---especially the vectorlike quarks---are somewhat stringently constrained by their non-observation at the LHC.

The most stringent limits on the $SU(2)_L$ singlet vectorlike quark with electric charge $2/3$, denoted by $U_1$ in the present framework, arise from its pair production via gluon fusion at the LHC, followed by decay into a $W$ boson and a $b$ quark. If one of the $W$ bosons decays leptonically and the other hadronically, the final state consists of exactly one charged lepton, missing transverse momentum, and jets. Assuming a branching ratio ${\rm Br}(U_1 \to W b)=1$, the latest ATLAS analysis \cite{ATLAS:2024gyc}, based on $140~\mathrm{fb}^{-1}$ of proton--proton collisions at $\sqrt{s}=13~\mathrm{TeV}$ excludes $m_{U_1} \lesssim 1.7~\mathrm{TeV}$ at $95\%$ confidence level (CL). A somewhat weaker limit, $m_{U_1} \lesssim 1.36~\mathrm{TeV}$, is obtained when the decay pattern ${\rm Br}(U_1 \to W b : h t : Z t) = 0.5 : 0.25 : 0.25$ is assumed. Similarly, a recent CMS analysis \cite{CMS:2025zwi} of single $U_1$ production followed by decay into $t$ and $h$ with a branching fraction of $25\%$ excludes $m_{U_1} < 2~\mathrm{TeV}$ at $95 \%$ CL. It can be seen from Table \ref{tab:VLFmass}, that these limits disfavour solution S1. The remaining two solutions lead to $m_{U_1}$ outside the kinematic reach of the present LHC.
\begin{table}[t]
    \centering
    \begin{tabular}{c c c c c}
    \toprule
        Mass & Limit & S1 & S2 & S3\\
        \midrule
       $m_{U_1}$& $\gtrsim {1.7} $ &$\textcolor{red}{\bf 0.51}$ & $2.46$ &$6.12$\\
       $m_{D_1}$ & $\gtrsim 1.54$  & $3.56$ & $17.6$ &$103.0$\\
       $m_{E_1}$&  $\gtrsim 0.4$  & $0.74$& $2.33$ & $5.36$\\
         \bottomrule
    \end{tabular}
    \caption{The lightest of vectorlike fermions in each sector for S1, S2 and S3 along with the current limits from the LHC. All the masses are in TeV.}
    \label{tab:VLFmass}
\end{table}

Analogous constraints apply to the $SU(2)_L$ singlet down-type vectorlike quarks $D_i$. The most stringent limits arise from pair production at the LHC, followed by decays into electroweak gauge bosons and third-generation quarks. The CMS Collaboration excludes masses $m_{D_1} \lesssim 1.56~\mathrm{TeV}$ assuming ${\rm Br}(D_1 \to W t)=1$ \cite{CMS:2022fck}, and $m_{D_1} \lesssim 1.54~\mathrm{TeV}$ for ${\rm Br}(D_1 \to Z b)=1$ \cite{CMS:2024xbc}, based on analyses of $138~\mathrm{fb}^{-1}$ of proton--proton collision data at $\sqrt{s}=13~\mathrm{TeV}$. A slightly weaker bound is obtained by ATLAS \cite{ATLAS:2022tla}, which excludes $m_{D_1} \lesssim 1.46~\mathrm{TeV}$ under similar branching-ratio assumptions. Single production of $D_1$ has also been investigated by ATLAS \cite{ATLAS:2023ixh}. In this case, $D_1$ masses up to $2~\mathrm{TeV}$ are excluded depending on their assumed couplings to $W$, $h$, and $Z$. As summarized in Table~\ref{tab:VLFmass}, all benchmark solutions considered in this work predict $D_1$ masses sufficiently heavy to evade the current direct-search constraints.

Unlike the case of vectorlike quarks, the current direct search limits on $SU(2)_L$ singlet vectorlike leptons are considerably weaker, owing to their much smaller production cross sections at the LHC. These states are pair-produced in proton--proton collisions via off-shell $\gamma$ or $Z$ exchange and subsequently decay into $W$, $Z$, or $h$ in association with a neutrino or a charged lepton. Assuming branching ratios of $50\%$, $25\%$, and $25\%$ into these channels, respectively, masses up to $m_{E_1} \lesssim 0.4~\mathrm{TeV}$ are excluded at $95\%$ CL \cite{ATLAS:2024mrr}. This bound is significantly weaker and does not impose any meaningful constraint on the spectra obtained for the benchmark solutions considered here, as can be seen from Table~\ref{tab:VLFmass}.

\subsection{Neutral meson-antimeson transitions}
\label{subsec:meson_transitions}
The flavour-violating couplings of the $X$ and $Z$ bosons contribute to the neutral meson–antimeson oscillations at the tree level. To quantify these  effects, we adopt an operator-based analysis and derive the corresponding effective Wilson coefficients (WCs) within our framework. The procedure is similar to the one earlier used and described at length in \cite{Mohanta:2022seo}. The WCs relevant for $\Delta S = 2$ neutral $M$–$\overline{M}$ transitions ($M=D^0$, $K^0$, $B_d^0$, $B_s^0$) can be obtained from the effective Hamiltonian,
\be \label{H_MMbar}
H_{\rm eff} = \sum_{i=1}^{5} C^i_M\, Q_i + \sum_{i=1}^{3} \tilde{C}^i_M\, \tilde{Q}_i\,,
\ee
where the explicit forms of the dimension-six operators $Q_i$ and $\tilde{Q}_i$ are given in \cite{UTfit:2007eik,Ciuchini:1998ix}.

The couplings of $X$-boson with the SM fermions are derived in Eq. (\ref{dgX}). When the former is integrated out, it induces the following WCs at the scale $\mu=M_X$ \cite{Smolkovic:2019jow},
\beqa \label{WCs_X}
C_D^{1\,(X)} &=& \frac{1}{2M_X^2}\left[\left(g^X_{uL}\right)_{12}\right]^2, ~~
\tilde{C}_D^{1\,(X)} = \frac{1}{2M_X^2}\left[\left(g^X_{uR}\right)_{12}\right]^2, ~~
C_D^{5\,(X)} = \frac{-2}{M_X^2}\left(g^X_{uL}\right)_{12}\left(g^X_{uR}\right)_{12}, \nonumber \\ 
C_K^{1\,(X)} &=& \frac{1}{2M_X^2}\left[\left(g^X_{dL}\right)_{12}\right]^2, ~~
\tilde{C}_K^{1\,(X)} = \frac{1}{2M_X^2}\left[\left(g^X_{dR}\right)_{12}\right]^2, ~~
C_K^{5\,(X)} = \frac{-2}{M_X^2}\left(g^X_{dL}\right)_{12}\left(g^X_{dR}\right)_{12}, \nonumber\\
C_{B_d}^{1\,(X)} &=& \frac{1}{2M_X^2}\left[\left(g^X_{dL}\right)_{13}\right]^2, ~~
\tilde{C}_{B_d}^{1\,(X)} = \frac{1}{2M_X^2}\left[\left(g^X_{dR}\right)_{13}\right]^2, ~~
C_{B_d}^{5\,(X)} = \frac{-2}{M_X^2}\left(g^X_{dL}\right)_{13}\left(g^X_{dR}\right)_{13}, \nonumber\\
C_{B_s}^{1\,(X)} &=& \frac{1}{2M_X^2}\left[\left(g^X_{dL}\right)_{23}\right]^2, ~~
\tilde{C}_{B_s}^{1\,(X)} = \frac{1}{2M_X^2}\left[\left(g^X_{dR}\right)_{23}\right]^2,~~
C_{B_s}^{5\,(X)} = \frac{-2}{M_X^2}\left(g^X_{dL}\right)_{23}\left(g^X_{dR}\right)_{23}.
\eeqa
The remaining WCs are zero at this scale due to the vectorial nature of $X$-boson mediated current. The modification to $Z$-boson couplings to the SM fermions also gives rise to the tree-level FCNC couplings as discussed in Section \ref{subsec:BSMZ}. The effects are parametrised in terms of the off-diagonal elements of $\delta g^Z_{fL}$ in Eq. (\ref{dgZ}). Upon integrating our the $Z$-boson, the non-vanishing WCs generated at the scale $\mu=M_Z$ are identified as, 
\beqa \label{WCs_Z}
C_{D}^{1\,(Z)} &=& \frac{g^2 c_\theta^2}{2c_W^2M_Z^2}\left[\left(\delta g^Z_{uL}\right)_{12}\right]^2,~~
C_{K}^{1\,(Z)} = \frac{g^2 c_\theta^2}{2c_W^2M_Z^2}\left[\left(\delta g^Z_{dL}\right)_{12}\right]^2,\nonumber \\
C_{B_d}^{1\,(Z)} &=& \frac{g^2 c_\theta^2}{2c_W^2M_Z^2}\left[\left(\delta g^Z_{dL}\right)_{13}\right]^2,~~
C_{B_s}^{1\,(Z)} = \frac{g^2 c_\theta^2}{2c_W^2M_Z^2}\left[\left(\delta g^Z_{dL}\right)_{23}\right]^2.
\eeqa
In comparision to Eq. (\ref{WCs_X}), the coefficients $\tilde{C}^1_M$ and $C^5_M$ are zero since $\delta g^Z_{fR}=0$ at the leading order.

Since both $g^{X}_{fL,fR}$ and $\delta g^{Z}_{dL}$ arise at $\mathcal{O}(\rho^{2})$ and are proportional to $\Omega_{fL,fR}$, the $Z$-boson--mediated FCNCs generically dominate over those induced by the $X$ boson, owing to the hierarchy $M_Z \ll M_X$ in the present framework. This behaviour is in sharp contrast to conventional radiative mass--generation models, in which the $X$ boson typically features $\mathcal{O}(1)$ couplings and hence provides the dominant source of FCNC effects. Furthermore, as discussed in Section~\ref{subsec:BSMFCNC} and illustrated by the benchmark solutions, the off-diagonal entries of $\Omega_{uL}$ and $\Omega_{uR}$ are parametrically larger than those in the other fermion sectors. One therefore expects the leading FCNC effects to arise in $D^{0}$-$\overline{D}^{0}$ mixing.
\begin{table}[t]
\centering
\setlength{\tabcolsep}{7pt}
\begin{tabular}{c c c c c}
\toprule
WC
& Limit
& S1
& S2
& S3\\
\midrule
$|{C_D^{1}}|$                    
& $< 7.2\times 10^{-13}$ 
& $\textcolor{red}{\bf 2.6\times 10^{-11}}$
& $5.9\times 10^{-14}$ 
& $6.0\times 10^{-15}$ \\

$|\tilde{C}_D^{1}|$            
& $< 7.2\times 10^{-13}$ 
& $4.7\times 10^{-14}$ 
& $4.6\times 10^{-15}$ 
& $5.7\times 10^{-16}$ \\

$|C_D^{4}|$                    
& $< 4.8\times 10^{-14}$ 
& $\textcolor{red}{\bf 1.3\times 10^{-12}}$
& $5.3\times 10^{-15}$ 
& $3.3\times 10^{-16}$ \\

$|C_D^{5}|$                    
& $< 4.8\times 10^{-13}$ 
& $\textcolor{red}{\bf 1.1\times 10^{-12}}$
& $3.6\times 10^{-15}$ 
& $2.0\times 10^{-16}$ \\
\midrule

$\mathrm{Re}\,{ C_K^{1}}$         
& $[-9.6,\,9.6]\times 10^{-13}$ 
& $2.1\times 10^{-15}$ 
& $2.9\times 10^{-17}$ 
& $-1.5\times 10^{-20}$ \\

$\mathrm{Re}\,\tilde{C}_K^{1}$ 
& $[-9.6,\,9.6]\times 10^{-13}$ 
& $2.8\times 10^{-14}$ 
& $1.0\times 10^{-16}$ 
& $-6.0\times 10^{-17}$ \\

$\mathrm{Re}\,C_K^{4}$         
& $[-3.6,\,3.6]\times 10^{-15}$ 
& $\textcolor{red}{\bf 1.6\times 10^{-14}}$
& $3.5\times 10^{-17}$ 
& $-2.9\times 10^{-19}$ \\

$\mathrm{Re}\,C_K^{5}$         
& $[-1.0,\,1.0]\times 10^{-14}$ 
& $1.0\times 10^{-14}$ 
& $1.8\times 10^{-17}$ 
& $-1.4\times 10^{-19}$ \\
\addlinespace

$\mathrm{Im}\,{ C_K^{1}}$         
& $[-9.6,\,9.6]\times 10^{-13}$ 
& $1.7\times 10^{-15}$ 
& $3.2\times 10^{-17}$ 
& $1.5\times 10^{-20}$ \\

$\mathrm{Im}\,\tilde{C}_K^{1}$ 
& $[-9.6,\,9.6]\times 10^{-13}$ 
& $-2.0\times 10^{-14}$ 
& $-1.2\times 10^{-16}$ 
& $-4.2\times 10^{-17}$ \\

$\mathrm{Im}\,C_K^{4}$         
& $[-1.8,\,0.9]\times 10^{-17}$ 
& $\textcolor{red}{\bf 3.3\times 10^{-16}}$
& $-5.8\times 10^{-19}$ 
& $2.5\times 10^{-20}$ \\

$\mathrm{Im}\,C_K^{5}$         
& $[-1.0,\,1.0]\times 10^{-14}$ 
& $2.2\times 10^{-16}$ 
& $-3.0\times 10^{-19}$ 
& $1.2\times 10^{-20}$ \\
\midrule

$|{C_{B_d}^{1}}|$                
& $< 2.3\times 10^{-11}$ 
& $1.4\times 10^{-16}$ 
& $3.0\times 10^{-18}$ 
& $3.4\times 10^{-21}$ \\

$|\tilde{C}_{B_d}^{1}|$        
& $< 2.3\times 10^{-11}$ 
& $5.1\times 10^{-15}$ 
& $1.5\times 10^{-17}$ 
& $6.8\times 10^{-17}$ \\

$|C_{B_d}^{4}|$                
& $< 2.1\times 10^{-13}$ 
& $6.8\times 10^{-16}$ 
& $1.5\times 10^{-18}$ 
& $6.0\times 10^{-20}$ \\

$|C_{B_d}^{5}|$                
& $< 6.0\times 10^{-13}$ 
& $8.4\times 10^{-16}$ 
& $1.4\times 10^{-18}$ 
& $5.0\times 10^{-20}$ \\
\midrule

$|{C_{B_s}^{1}}|$                
& $< 1.1\times 10^{-9}$ 
& $1.7\times 10^{-15}$ 
& $3.6\times 10^{-17}$ 
& $5.2\times 10^{-20}$ \\

$|\tilde{C}_{B_s}^{1}|$        
& $< 1.1\times 10^{-9}$ 
& $3.9\times 10^{-14}$ 
& $2.6\times 10^{-16}$ 
& $4.3\times 10^{-16}$ \\

$|C_{B_s}^{4}|$                
& $< 1.6\times 10^{-11}$ 
& $6.8\times 10^{-15}$ 
& $2.2\times 10^{-17}$ 
& $5.9\times 10^{-19}$ \\

$|C_{B_s}^{5}|$                
& $< 4.5\times 10^{-11}$ 
& $8.3\times 10^{-15}$ 
& $2.0\times 10^{-17}$ 
& $5.0\times 10^{-19}$ \\
\bottomrule
\end{tabular}
\caption{Estimated values of various Wilson coefficients (in ${\rm GeV}^{-2}$) obtained for three benchmark solutions and corresponding experimental limits at $95\%$ CL. The excluded ones are shown in red.}
\label{tab:WC_constraints}
\end{table}

To compare with the present experimental limits on WCs, we evolve $C_D^{i\,(Z,X)}$ from $\mu=M_{Z,X}$ to $\mu=2.8$ GeV using the RGE equations given in \cite{UTfit:2007eik} and combine them to estimate the total contribution $C_D^{i}$. It is seen that the running induces appreciable magnitude for $C_D^4$ at low energy, while the other vanishing WCs do not get generated. A similar exercise is carried out for $C_K^{i\,(Z,X)}$ which are evolved to $\mu=2$ GeV using the RGE equations listed in \cite{Ciuchini:1998ix}. The coefficients $C_{B_{d,s}}^{i\,(Z,X)}$ are run down to $\mu=4.6$ GeV following \cite{Becirevic:2001jj}. The  values of these coefficients computed in this way are listed in Table \ref{tab:WC_constraints} for three benchmark solutions, and they are compared with the corresponding present experimental limits extracted from \cite{UTfit:2007eik,Aebischer:2020dsw}.

It is observed that the solution with $M_X = 1$ TeV is excluded by both $D^0$-$\overline{D}^0$ and $K^0$-$\overline{K}^0$ mixing constraints. The dominant contribution arises from $C^{1}_{D}$, which is driven by sizeable unitarity violation in the up-type quark sector, together with the enhanced mediation via the $Z$ boson. The WCs $C^{4,5}_{D}$, which are primarily induced by $X$-boson exchange, are also found to be sizeable. In contrast, the WCs in the down-type quark sector are comparatively suppressed, as shown in Table~\ref{tab:WC_constraints}. Nevertheless, the stringent limits from $K^0$-$\overline{K}^0$ mixing are sufficient to exclude the benchmark point S1. The magnitudes of flavour violation involving the first-second and second-third generations are found to be comparable. However, the current experimental bounds are considerably weaker in the latter case. It is further observed that the benchmark solutions S2 and S3 remain consistent with present constraints. Nonetheless, an improvement of one to two orders of magnitude in the experimental sensitivity to $D^0$-$\overline{D}^0$ mixing would have a significant impact on their viability.

\subsection{Rare decays of top quark}
\label{subsec:rare_top}
Relatively large flavour violation in the up-type quark sector motivates an investigation of rare top decays within the present framework. The decays $t \to Z\, c(u)$ and $t \to h\, c(u)$ can occur already at tree level, unlike in the SM. Processes such as $t \to \gamma\, c(u)$ and $t \to g\, c(u)$ become operative at the 1-loop level. Within the SM, these are mediated by 1-loop penguin diagrams with down-type quarks and the $W$ boson inside the loop and are strongly suppressed by the GIM mechanism. In the present model, new contributions can arise through $Z$, $h$, or $X$ bosons and quarks of charge $2/3$ running in the loop.

The partial widths of two-body tree-level decays of the top quark are estimated as
\be \label{br:tZHq}
\Gamma\left(t \to Z q_a, h q_a\right) = \frac{1}{16 \pi m_t^3}\,\left[(m_t^2-m^2_{q_a}-m_{Z,h}^2)^2-4m_{q_a}^2m_{Z,h}^2\right]^{\frac{1}{2}}\,\left| {\cal M}^{Z,h}_{q_a} \right|^2\,,\ee
where $a=1,2$ corresponds to $q_1=u$ and $q_2=c$, and ${\cal M}$ denotes the relevant matrix elements summed over all helicities and polarisations of the initial and final states \cite{Goodsell:2017pdq}. Using Eqs. (\ref{dgZ}) and (\ref{dgh}), the matrix elements are computed as
\beqa \label{ME}
\left| {\cal M}^Z_{q_a} \right|^2&=& \frac{g^2 c_\theta^2}{2 c_W^2M_Z^2}\left(m_t^4+m_{q_a}^4+m_{q_a}^2 M_Z^2-2M_Z^4+m_t^2(-2m_{q_a}^2+M_Z^2)\right)\,\left| \left(\delta g_{u_L}^Z \right)_{a 3} \right|^2\,, \nonumber \\
\left| {\cal M}^h_{q_a} \right|^2&=& \frac{1}{2} \left(m_t^2+m_{q_a}^2-m_h^2\right)\, \left| \left(\delta g^h_u\right)_{a3} \right|^2\,, \eeqa
within the present framework.

For $t \to \gamma\, c(u)$, which arises at the 1-loop level, the estimation is carried out following \cite{Aguilar-Saavedra:2002lwv}. In general, the transition amplitude for $t \to \gamma\, q_a$ can be parametrised in terms of the form factors $A^\gamma_{q_a}$ and $B^\gamma_{q_a}$ as
\be \label{MA:tcg}
{\cal M}(t \to \gamma\, q_a) = \bar{u}(p_{q_a})\, \left(i \sigma^{\mu \nu} p^\prime_\nu \left(A^\gamma_{q_a} + B^\gamma_{q_a} \gamma_5 \right) \right)\,u(p_{t})\,\epsilon^*_\mu(p^\prime)\,, \ee
where $p_{q_a}$ is the momentum of $q_a$ and $p^\prime = p_t - p_{q_a}$. An analogous amplitude exists for $t \to g\, q_a$ with form factors $A^g_{q_a}$ and $B^g_{q_a}$. The partial widths of these processes are then given by
\beqa \label{gamma:tcg}
\Gamma(t \to q_a\, \gamma/g) = \frac{C_F}{\pi} \left(\frac{m_t^2-m_c^2}{2 m_t} \right)^3\, \left(\left| A^{\gamma,g}_{q_a} \right|^2 + \left| B^{\gamma,g}_{q_a} \right|^2 \right)\,,
\eeqa
where $C_F = 1$ for $t \to q_a\,\gamma$ and $C_F = 4/3$ for $t \to q_a\, g$.

As mentioned before, new contributions to these processes arise through diagrams involving $Z$, $h$, or $X$ bosons in the loop. Since flavour violation from all of these sources appears at the same ${\cal O}(\rho^2)$, we consider only the $Z$-boson mediated contributions to estimate the size of these processes. The $X$-boson contribution is suppressed due to $M_X \gg M_Z$, while flavour-violating couplings involving $h$ are further suppressed by an additional factor of ${\cal O}(m_f/v)$ compared to those involving the $Z$ boson. The form factors computed for the $Z$-mediated diagrams for both processes are given in \cite{Aguilar-Saavedra:2002lwv}, and for convenience we reproduce them in Appendix \ref{app:Formfactors} in terms of $\delta g^Z_{uL}$.
\begin{table}[t]
\centering
\setlength{\tabcolsep}{5pt}
\begin{tabular}{c c c c c c c}
\toprule
Observable
& Current limit
& HL-LHC limit
& S1
& S2
& S3 \\
\midrule
$\mathrm{Br}(t \to Z c)$
& $2.4{\times}10^{-4}$
& $2.3{\times}10^{-5}$
& $1.8{\times}10^{-5}$
& $5.4{\times}10^{-8}$
& $5.0{\times}10^{-9}$ \\

$\mathrm{Br}(t \to Z u)$
& $1.7{\times}10^{-4}$
& $7.3{\times}10^{-6}$
& $2.0{\times}10^{-7}$
& $9.3{\times}10^{-10}$
& $6.0{\times}10^{-11}$ \\
\addlinespace

$\mathrm{Br}(t \to h c)$
& $7.3{\times}10^{-4}$
& $8.5{\times}10^{-5}$
& $4.4{\times}10^{-7}$
& $1.3{\times}10^{-9}$
& $1.2{\times}10^{-10}$ \\

$\mathrm{Br}(t \to h u)$
& $1.9{\times}10^{-4}$
& $8.5{\times}10^{-5}$
& $4.9{\times}10^{-9}$
& $2.2{\times}10^{-11}$
& $1.5{\times}10^{-12}$ \\
\addlinespace
$\mathrm{Br}(t \to c\gamma)$
& $4.0{\times}10^{-4}$
& $5.2{\times}10^{-5}$
& $2.8{\times}10^{-11}$
& $1.5{\times}10^{-13}$
& $6.0{\times}10^{-15}$ \\

$\mathrm{Br}(t \to u\gamma)$
& $8.9{\times}10^{-5}$
& $6.1{\times}10^{-6}$
& $1.3{\times}10^{-12}$
& $1.1{\times}10^{-15}$
& $1.3{\times}10^{-16}$ \\
\addlinespace
$\mathrm{Br}(t \to cg)$
& $4.1{\times}10^{-4}$
& $5.2{\times}10^{-5}$
& $1.2{\times}10^{-9}$
& $6.3{\times}10^{-12}$
& $2.5{\times}10^{-13}$ \\

$\mathrm{Br}(t \to ug)$
& $2.0{\times}10^{-5}$
& $6.1{\times}10^{-6}$
& $5.3{\times}10^{-11}$
& $4.4{\times}10^{-14}$
& $5.5{\times}10^{-15}$ \\
\bottomrule
\end{tabular}
\caption{Predictions for branching ratios of rare top quark decays for the benchmark solutions. The current limits are taken from \cite{dEnterria:2023wjq} while the HL-LHC sensitivities are taken from \cite{Liu:2020bem}, \cite{ATLAS:2016qxw} and \cite{Cerri:2018ypt} for the decays $t \to Z q_a$, $t \to h q_a$ and $t \to q_a \gamma (g)$, respectively.}
\label{tab:top_fcnc}
\end{table}

We compute the branching ratios of all the above decays numerically for the three solutions listed in Table \ref{tab:fit_results}. For the external top quark, we use its pole mass $m_t = 172.56$ GeV \cite{ParticleDataGroup:2024cfk}. We also neglect RGE effects in $\delta g^Z_{uL}$ and $\delta g^h_{u}$ for these estimates, as such effects are not significant enough to affect the order-of-magnitude results. The results are displayed in Table \ref{tab:top_fcnc}. It can be noticed that the limits provided by the rare decays of the top quark on the present framework are not as competitive as the ones coming from $D$-$\overline{D}$ oscillations. All the solutions including S1 lead to the branching ratios well within the present limits. Among all the observables, ${\rm Br}(t \to Z c)$ comes closest to the present limit. The branching fractions of loop-induced decays of top are several orders of magnitude smaller than the tree-level decays. As can be seen from the Table \ref{tab:top_fcnc}, the high luminosity upgrade of the LHC may also not be able to exclude these predictions. 

Comparing these results with the SM predictions for the same observables, we find that all solutions lead to significantly enhanced branching fractions for the tree-level decays. For instance, a leading-order SM calculation yields ${\rm Br}(t \to Z c) \simeq 1.0{\times}10^{-14}$, ${\rm Br}(t \to Z u) \simeq 8.0{\times}10^{-17}$ \cite{Aguilar-Saavedra:2004mfd} while ${\rm Br}(t \to h c) \simeq 4.19{\times}10^{-15}$, ${\rm Br}(t \to h u) \simeq 3.66{\times}10^{-17}$ \cite{Altmannshofer:2019ogm}. For the top decaying into gluon or photon, the SM predictions stand as 
${\rm Br}(t \to c g) \simeq 5.31{\times}10^{-12}$, ${\rm Br}(t \to u g) \simeq 3.81{\times}10^{-14}$, ${\rm Br}(t \to c \gamma) \simeq 4.55{\times}10^{-14}$ and ${\rm Br}(t \to u \gamma) \simeq 3.26{\times}10^{-16}$ \cite{Balaji:2020qjg}. The corresponding predictions in scenarios S2 and S3 are of comparable magnitude, and in some cases even smaller.

\subsection{Rare decays of mesons and Higgs}
\label{subsec:rare_meson}
We now discuss flavour-violating decays of some relevant mesons and Higgs for completeness, although they do not provide noteworthy constraint on the present model due to relatively suppressed effects in these sectors.   

The neutral-current transitions in the down-type quark sector of the type
$b \to s \ell^+ \ell^-$ mediated by the flavour-violating couplings of the $Z$ boson contribute to the decay $B_s \to \mu^+ \mu^-$ at the tree-level itself. Such contributions are parametrised by effective Wilson coefficients  $C_9$ and $C_{10}$. The deviations from their SM values, $C^{\rm SM}_9 = 4.32$ and $C^{\rm SM}_{10} = -4.41$ \cite{Bobeth:2003at}, can be parameterised following \cite{Alonso:2018bcg} as
\be
\delta C_{10}
= -\frac{1}{1 - 4 s_{\theta_W}^2}\,\delta C_9
= -\frac{2 \pi}{\alpha\, V^{\rm CKM *}_{32} V^{\rm CKM}_{33}}
\left(\delta g^Z_{dL}\right)_{23}\,.
\ee
The current global fit to $b \to s \mu^+ \mu^-$ data constrain these deviations to $\delta C_9^{\rm b.f.} = -0.88 \pm 0.16$ and $\delta C_{10}^{\rm b.f.} = 0.09 \pm 0.19$ \cite{Hurth:2025vfx}. A negative non-zero value for the first is preferred, while the second is consistent with zero. For the benchmark solution S1, we obtain $\delta C_9 \simeq -0.009-0.006\,i$ and $\delta C_{10} \simeq -0.121-0.080\,i$, which are within the allowed ranges. The corresponding contributions for S2 and S3 are even smaller. Note that we have neglected the RGE effects while making these comparisons, however, they are not expected to change the results by order of magnitudes. 

Analogously, the measurement of the rare kaon decay $K^+ \to \pi^+ \nu \bar{\nu}$~\cite{E949:2008btt} constrains the Wilson coefficients governing the $s \to d \nu \bar{\nu}$ transition, yielding $\delta C_\nu = 2.7^{+2.2}_{-3.2}$~\cite{Alonso:2018bcg}. Within the same parameterisation, the prediction for solution S1 is $\delta C_\nu \simeq -13.8435+8.0039\, i$, corresponding to a sizeable deviation from the experimental  value. However, the predictions for solutions S2 and S3, $\delta C_\nu \simeq -1.867+0.960 \,i$ and $\delta C_\nu \simeq -0.0468+0.0046\, i$, respectively, stay compatible with current bounds.

The FCNC effects arising from modifications of Higgs couplings can be probed by comparing theoretical predictions with the ATLAS constraints obtained from searches for lepton-flavour-violating Higgs decays into $\mu\tau$ and $e \tau$ final states \cite{ATLAS:2023mvd}. The corresponding experimental limits translate into the bounds $\sqrt{|\left(\delta g^h_l\right)_{23}|^2 + |\left(\delta g^h_l\right)_{32}|^2} < 1.2 \times 10^{-3}$ and $\sqrt{|\left(\delta g^h_l\right)_{13}|^2 + |\left(\delta g^h_l\right)_{31}|^2} < 1.4 \times 10^{-3}$, respectively. For the benchmark solution S1, the predicted values are $2.4 \times 10^{-8}$ and $4.3 \times 10^{-8}$, respectively. They are already several orders of magnitude below the current experimental sensitivity, implying that the modifications to Higgs couplings are not presently severely constraining the present setup.

\subsection{Lepton flavour violation}
Since the basic mechanism is extended to the charged leptons, flavour violation in this sector also becomes unavoidable. As before, such effects are induced through $Z$, $X$, or $h$ mediation, and we discuss only the $Z$-induced contributions for the reasons already mentioned in the previous subsections. Deviations from the unitarity in the couplings of leptons with the $Z$ boson, as quantified in Eq. (\ref{dgZ}), induce $\mu \to e$ conversion in nuclei and trilepton decays already at tree level. Likewise, the transitions $\ell_\alpha \to \ell_\beta \gamma$ are generated at the 1-loop level.

The muon can transition into an electron in the field of a nucleus by exchanging a $Z$ boson with the nucleus. The most stringent limit on such a process is presently obtained from the SINDRUM II experiment \cite{SINDRUMII:2006dvw}, which used an $^{197}$Au nucleus. The branching ratio of this process is computed as
\be \label{Br:mu2e}
{\rm Br}(\mu \to e) = \frac{2 G_F^2}{\omega_{\rm capt}}\, (V^{(p)})^2\,\left(\left|g_{LV}^{(p)} \right|^2 + \left|g_{RV}^{(p)} \right|^2 \right)\,,\ee
where $\omega_{\rm capt}$ is the muon capture rate, and $V^{(p)}$ is a proton distribution integral for the given nucleus. The couplings are parametrized as
\be \label{g1:mu2e}
g_{LV,RV}^{(p)} = 2 g_{LV,RV}^{(u)} + g_{LV,RV}^{(d)}\,, \ee
and the flavour-violating $Z$-boson interactions are given by
\be \label{g2:mu2e}
g^{(f)}_{LV} = 4 c_\theta^2\, \left(T_3^f - 2 s_W^2 Q_f \right)\, \left(\delta g^Z_{eL} \right)_{12}\,,~~g^{(f)}_{RV} = 0\,,\ee
for $f=u,d$. Using $\left(\delta g^Z_{eL} \right)_{12}$ computed from the three benchmark solutions and $V^{(p)}=0.0974\, m_{\mu}^{5/2}$, $\omega_{\rm capt} = 13.07 \times 10^{6}\,{\rm sec}^{-1}$, we estimate ${\rm Br}(\mu \to e)$. The results are listed in Table \ref{tab:lfv_predictions} along with the present experimental upper limit.
\begin{table}[t]
\centering
\begin{tabular}{c c c c c}
\toprule
Observable 
& Limit
& S1
& S2
& S3 \\
\midrule
$\mathrm{Br}(\mu \to e)_{\rm Au}$               
& $7.0 \times 10^{-13}$ 
& $\textcolor{red}{\bf 9.8 \times 10^{-12}}$
& $1.4 \times 10^{-14}$ 
& $9.0 \times 10^{-15}$ \\

\addlinespace
$\mathrm{Br}(\mu \to 3e)$            
& $1.0 \times 10^{-12}$ 
& $\textcolor{red}{\bf 1.6 \times 10^{-10}}$
& $2.2 \times 10^{-13}$ 
& $7.8 \times 10^{-14}$ \\

$\mathrm{Br}(\tau \to 3\mu)$         
& $2.1 \times 10^{-8}$  
& $3.6 \times 10^{-10}$  
& $5.4 \times 10^{-13}$ 
& $4.4 \times 10^{-13}$ \\

$\mathrm{Br}(\tau \to 3e)$           
& $2.7 \times 10^{-8}$  
& $3.5 \times 10^{-12}$ 
& $4.1 \times 10^{-15}$ 
& $3.8 \times 10^{-15}$ \\

\addlinespace
$\mathrm{Br}(\mu \to e\gamma)$       
& $4.2 \times 10^{-13}$ 
& $\textcolor{red}{\bf 7.9 \times 10^{-11}}$
& $1.1 \times 10^{-13}$ 
& $3.9 \times 10^{-15}$ \\

$\mathrm{Br}(\tau \to \mu\gamma)$    
& $4.4 \times 10^{-8}$  
& $2.9 \times 10^{-14}$ 
& $3.4 \times 10^{-17}$ 
& $3.7 \times 10^{-17}$ \\

$\mathrm{Br}(\tau \to e\gamma)$      
& $3.3 \times 10^{-8}$  
& $3.0 \times 10^{-16}$ 
& $2.3 \times 10^{-19}$ 
& $1.8 \times 10^{-19}$ \\
\bottomrule
\end{tabular}
\caption{Estimations of the charged lepton flavour violating processes for three benchmark solutions and the present experimental upper limits from \cite{Calibbi:2017uvl}. The numbers shown in red exceed the corresponding experimental limits.}
\label{tab:lfv_predictions}
\end{table}

The partial decay widths of $\ell_\alpha \to 3 \ell_\beta$ are computed following \cite{Smolkovic:2019jow,Heeck:2016xkh}. At the leading order in the charged-lepton masses, they are estimated as,
\beqa \label{Gamma:lto3l}
\Gamma(\ell_\alpha \to 3 \ell_\beta) &=& \frac{m_{l_\alpha}^5}{768 \pi^3 M_Z^4}\, \Big[ 4\,{\rm Re}\left( (X_{eV})_{\beta \alpha} (X_{eA})_{\beta \alpha} (X_{eV})^*_{\beta \beta} (X_{eA})^*_{\beta \beta} \right) \Big. \nonumber \\
&+& 3 \left(\left|(X_{eV})_{\beta \alpha} \right|^2 + \left|(X_{eA})_{\beta \alpha} \right|^2 \right)  \left(\left|(X_{eV})_{\beta \beta} \right|^2 + \left|(X_{eA})_{\beta \beta} \right|^2 \right) \Big]\,.
\eeqa
Here, 
\be \label{X_VA}
X_{eV,eA} = \frac{1}{2} \left(X_{eL} \pm X_{eR} \right)\,,\ee
and
\beqa \label{X_eLeR}
\left(X_{eL}\right)_{\beta \alpha} = \frac{g c_\theta}{c_W} \left(\left(T^e_3 - Q_e s_W^2\right)\, \delta_{\beta \alpha} + \left(\delta g_{eL}^Z \right)_{\beta \alpha}\right), ~~\left(X_{eR}\right)_{\beta \alpha} = - \frac{g c_\theta}{c_W}\, Q_e s_W^2\, \delta_{\beta \alpha}\,. 
\eeqa
The estimated branching ratios for the benchmark solutions are given in Table \ref{tab:lfv_predictions} together with the corresponding experimental upper limit from \cite{Calibbi:2017uvl}. 

Finally, the 1-loop process $\ell_\alpha \to \ell_\beta \gamma$ can be estimated in an identical manner to that discussed for $t \to \gamma\, c(u)$ in the previous subsection. The expression for the partial width, Eq. (\ref{gamma:tcg}), and the form factors listed in Appendix \ref{app:Formfactors} can be straightforwardly modified by replacing the masses and couplings appropriately. The estimates for ${\rm Br}(\mu \to e \gamma)$, ${\rm Br}(\tau \to \mu \gamma)$, and ${\rm Br}(\tau \to e \gamma)$ carried out in this way are listed in Table \ref{tab:lfv_predictions}.

It can be seen that the predicted branching ratios for the benchmark scenario S1 are strongly constrained, with the processes $\mu$-$e$ conversion and $\mu \to 3 e$ exceeding the present limits by more than an order of magnitude. While the magnitudes of $\tau \to 3 \mu$ and $\tau \to 3e$ are almost similar, the latter is more stringently constrained by experiments. The branching ratios of radiative decays, $\ell_\alpha \to \ell_\beta \gamma$, are suppressed by one to four orders of magnitude compared to those of the tree-level processes. However, the rate of $\mu \to e \gamma$ predicted by S1 is large enough to be excluded by the present limit. The same by S2 stays close to the present limit and can be tested with if the corresponding experimental limit improves by an order of magnitude. For the remaining processes, the predicted rates by S2 and S3 are several orders of magnitude below the corresponding limits. Constraining these solutions through charged-lepton flavour violation therefore requires significant improvements in the present experimental limits. It is noteworthy that some of the flavour violation in the charged-lepton sector is sizeable enough to be constrained by the present limits, whereas the same in the down-type quark sector is negligible. This is also due to the fact that, among the spectrum of heavy fermions, the charged-lepton states are lighter than the down-type quarks, as can be seen from the spectrum shown in Fig. \ref{fig4}.

\subsection{Low-energy and electroweak precision observables}
So far, our discussion has focused on the flavour-violating couplings of the $Z$ boson, which are strongly constrained by rare decays. The shifts in the flavour-conserving couplings are also severely constrained by several low-energy observables and electroweak precision data. An extensive analysis of these effects has been carried out in \cite{Falkowski:2017pss}, which performs a global fit to these shifts and to the Wilson coefficients of dimension-6 flavour-conserving operators using a total of 264 experimental inputs. The latter include precision measurements of $Z$ decays, parity violation in atoms and in scattering experiments, semileptonic and leptonic decays of hadrons, quark-pair production, and forward–backward asymmetries, among other observables. Our parametrisation of the modifications in the $Z$-boson couplings given in Section \ref{subsec:BSMZ} is identical to that used in \cite{Falkowski:2017pss}. Therefore, it is straightforward to compare our estimates with the results of the global fit. 
\begin{table}[t]
\centering
\begin{tabular}{ccccc}
\toprule
Couplings 
& Constraint 
& S1 
& S2
& S3 \\
\midrule
$(\delta g^Z_{uL})_{11}$ 
& $(-0.8 \pm 3.1) \times 10^{-2}$ 
& $-7.4\times 10^{-4}$ 
& $-7.4\times 10^{-5}$ 
& $-1.6\times 10^{-5}$ \\

$(\delta g^Z_{uL})_{22}$ 
& $(-0.15  \pm 0.36) \times 10^{-2}$ 
& $\textcolor{red}{\bf -2.2\times 10^{-2}}$
& $-1.3\times 10^{-3}$ 
& $-2.9\times 10^{-4}$ \\

$(\delta g^Z_{uL})_{33}$ 
& $(-0.3 \pm 3.8) \times 10^{-2}$ 
& $-2.3\times 10^{-2}$ 
& $-1.3\times 10^{-3}$ 
& $-3.2\times 10^{-4}$ \\
\addlinespace
$(\delta g^Z_{dL})_{11}$ 
& $(-0.9 \pm 4.4) \times 10^{-2}$ 
& $4.2\times 10^{-6}$ 
& $6.3\times 10^{-7}$ 
& $1.3\times 10^{-8}$ \\

$(\delta g^Z_{dL})_{22}$ 
& $(0.9  \pm 2.8) \times 10^{-2}$ 
& $4.0\times 10^{-5}$ 
& $5.1\times 10^{-6}$ 
& $1.2\times 10^{-7}$ \\

$(\delta g^Z_{dL})_{33}$ 
& $(0.33 \pm 0.17) \times 10^{-2}$ 
& {$4.2\times 10^{-5}$} 
& $5.3\times 10^{-6}$ 
& $1.4\times 10^{-7}$ \\
\addlinespace
$(\delta g^Z_{eL})_{11}$ 
& $(-2.3 \pm 2.8) \times 10^{-4}$ 
& $1.3\times 10^{-6}$ 
& $4.5\times 10^{-8}$ 
& $4.5\times 10^{-8}$ \\

$(\delta g^Z_{eL})_{22}$ 
& $(0.1 \pm 1.2) \times 10^{-3}$ 
& $1.2\times 10^{-4}$ 
& $4.7\times 10^{-6}$ 
& $3.3\times 10^{-6}$ \\

$(\delta g^Z_{eL})_{33}$ 
& $(1.8 \pm 5.9) \times 10^{-4}$ 
& $1.5\times 10^{-4}$ 
& $6.0\times 10^{-6}$ 
& $4.8\times 10^{-6}$ \\
\bottomrule
\end{tabular}
\caption{Estimated values of corrections to the flavour conserving $Z$-boson couplings and comparison with the constraint from the global fit results \cite{Falkowski:2017pss}.}
\label{tab:Zcouplings}
\end{table}

We numerically compute all the deviations in $Z$-boson couplings for three benchmark solutions. As already shown in Section \ref{subsec:BSMZ}, only $\delta g^Z_{fL}$ are relevant since $\delta g^Z_{fR}$ are vanishing due to the nature of the extension. The comparison is presented in Table \ref{tab:Zcouplings}. It is seen that the deviations in the charged-lepton and down-type quark sectors are significantly small and stay well within the current ranges allowed by the global fit to electroweak precision data. The only deviation which is excluded by the current fit is in the $(2,2)$ element of $\delta g^Z_{uL}$. The predicted numbers in Table \ref{tab:Zcouplings} are in agreement with the anticipated pattern of the modification in $Z$ couplings across various sectors. 

The presence of mixing between $X$ and $Z$ bosons modify the $\rho$ parameter. At the leading order in the mixing angle $\theta$, the shift in the $\rho$ is given by,
\be \label{delta_rho}
\Delta \rho = \frac{g_X^2}{g^2 + g^{\prime 2}}\,\left(\frac{M_Z}{M_X}\right)^2\,. \ee
The latest results from the fit to electroweak precision observables allows $-6 \times 10^{-5} \leq \Delta \rho \leq 6.8 \times 10^{-4}$ at $95\%$ CL \cite{ParticleDataGroup:2024cfk}. This translates into $M_X/g_X \geq 4.6$ TeV. Hence, the benchmark solution S1 is also excluded from this consideration, while S2 and S3 are consistent with this constraint.

\section{Neutrino Masses}
\label{sec:nu_mass}
The model can be straightforwardly extended to accommodate nonvanishing neutrino masses in close analogy with the charged-lepton sector. This is achieved by introducing copies of SM-singlet but $U(1)_X$ charged fermion pairs $N_{L,R}$, together with three gauge singlet right-handed neutrinos $\nu_R$. Neutrino masses can then be generated in a way analogous to the charged fermion, which can be seen as a latticed-theory-space version of the so called Dirac-seesaw mechanism \cite{Roncadelli:1983ty}. Their smallness relative to the electroweak scale may be arranged by taking $M_N \gg M_{U,D,E}$. Nevertheless, given the possible Majorana nature of neutrinos and the comparatively mild hierarchy observed in the neutrino mass spectrum, the underlying mechanism responsible for neutrino mass generation could be qualitatively distinct from that of the charged fermions. We discuss one such possibility here as an example.

If the scale of lepton-number violation is assumed above the $U(1)_X$ breaking scale, the classic Weinberg operator, $\frac{1}{\Lambda} l_L l_L H H$, is forbidden, as it can be seen from Table \ref{tab:field}. The neutrino masses can then arise through dimension-7 operator, $\frac{1}{\Lambda^3} l_L l_L H H S^2$. This implies that $\Lambda$ scales as,
\be \label{}
\Lambda \approx 2 \times 10^8\,{\rm GeV}\,\times \left(\frac{v_S}{10\,{\rm TeV}}\right)^{2/3} \times \left(\frac{0.05\,{\rm eV}}{m_{\nu 3}}\right)^{1/3}\,.\ee
Consequently, the new physics responsible for neutrino mass generation can be at somewhat lower scale in comparison to the canonical seesaw mechanisms.  

\begin{figure}[t!]
\centering
\subfigure{\includegraphics[width=0.35\textwidth]{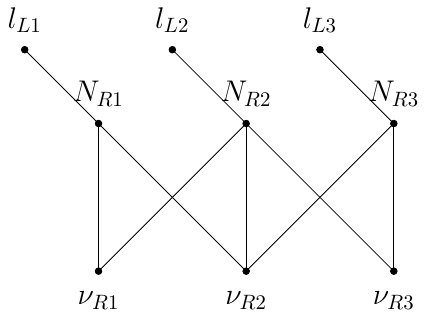}}
\caption{Interaction topology implementing the latticed version of  ``inverse-seesaw mechanism''. $\nu_{R \alpha}$ acquire Majorana masses while the mass terms involving $N_{R \alpha}$ and $\nu_{R \alpha}$ arise through $U(1)_X$ breaking.}
\label{fig5}
\end{figure}
Alternatively, a low-scale lepton number violation can also be realised. Consider for example three generations of the SM singlet fermions $\nu_{R \alpha}$ and $N_{R \alpha}$. Only the latter are charged under $U(1)_X$ with charges $+1$. The $\nu_{R \alpha}$ being complete singlets under the gauge symmetry can acquire Majorana masses which break lepton number by two units. Arranging structured interactions between these fermions as depicted in Fig. \ref{fig5}, the corresponding interaction terms are characterised by,
\be \label{Lnu}
-{\cal L}_\nu \supset  (y_\nu)_\alpha\, \overline{l_L}_{\alpha} \tilde{H} N_{R\alpha} + (y_\nu^\prime)_{\alpha \beta}\, \overline{N_R}_{\alpha}\,S^*\, \nu_{R \beta} + \frac{1}{2} \mu_{\alpha}\,\overline{\nu^c_R}_\alpha \nu_{R \alpha} + {\rm h.c.} \,.
\ee
Note that we choose only three sites in this case for the latticed theory space, since we intend to generate the masses for all three neutrinos at the leading order itself. The intergenerational mass hierarchy among the neutrinos is small and can be easily reproduced through ${\cal O}(1)$ couplings, as we show below.

After the $U(1)_X$ and electroweak symmetry breaking, the masses of neutral fermions can be written as 
\be \label{basis_nu}
-\frac{1}{2}  \left(\ba{ccc} \overline{\nu_L} & \overline{N^c_R} & \overline{\nu_R} \ea\right)\,{\cal M}_\nu\, \left( \ba{c} \nu_L^c \\ N_R \\ \nu_R^c \ea\right)  + {\rm h.c.} \,,
\ee
with a $9 \times 9$ Majornana mass matrix,
\be \label{calM_nu}
{\cal M}_\nu = \left(\ba{ccc} 0 & m_D & 0 \\
m_D^T & 0 & M_N\\
0 & M_N^T & \mu \ea\right)\,,\ee
and
\be \label{mD}
(m_D)_{\alpha \beta} = \frac{v}{\sqrt{2}}\, (y_\nu)_\alpha \delta_{\alpha \beta}\,,\quad \mu_{\alpha \beta} = \mu_\alpha\,\delta_{\alpha \beta}\,,\ee
and 
\be \label{MN}
(M_N)_{\alpha \beta} = \frac{v_S}{\sqrt{2}}\,(y^\prime_\nu)_{\alpha \beta} \equiv W_N\,\left[\,\delta_{\alpha \beta} + t_N\, (\delta_{\alpha+1,\beta} + \delta_{\alpha,\beta+1})\right]\,,\ee
The above can be seen as a specific realisation of the well-known ``inverse seesaw'' mechanism \cite{Mohapatra:1986bd,Abada:2014vea,CentellesChulia:2020dfh}. The block matrices have particular structures and the scale of $M_N$ is directly linked with the $U(1)_X$ breaking scale.

In the limit $\mu \ll m_D \ll M_N$, one gets the effective light neutrino mass matrix,
\be \label{M_nu}
M_\nu \simeq m_D\,(M_N^T)^{-1}\,\mu\,M_N^{-1}\,m_D^T\,.\ee
For $(y_{\nu})_\alpha \sim {\cal O}(1)$ and universal $\mu_\alpha \equiv \mu$, the masses of three light neutrinos are given by
\be \label{evn}
\frac{\mu v^2}{2 W_N^2}\,\left\{\frac{1}{\left(1-\sqrt{2}\, t_N\right)^2}\,,\,\,\frac{1}{\left(1+\sqrt{2}\, t_N \right)^2}\,,\,\, 1 \right\}\,.\ee
The above can easily reproduce the desired neutrino mass spectrum with ${\cal O}(1)$ values of $t_N$. For example, $t_N = -0.7726 + 1.8245\,i$ implies $m_{\nu 1}/m_{\nu 3} = 0.091$ and $m_{\nu 2}/m_{\nu 3}=0.15$ leading to the ratio of solar to atmospheric mass scales consistent with the present data. 

The parameter $\mu$ fixes the overall scale of neutrino masses, and it can be quantified as
\be \label{mu}
\mu \approx 165\,{\rm eV}\,\times \left(\frac{W_N}{10\,{\rm TeV}} \right)^2 \times \left(\frac{246\,{\rm GeV}}{v} \right)^2 \times \left(\frac{m_{\nu 3}}{0.05\,{\rm eV}} \right)\,.\ee
In the limit $\mu \to 0$, the lepton number gets restored as a symmetry of the model and hence $\mu \ll v,v_S$ is technically natural. Additional model building can also be carried out to justify small $\mu$ from some underlying high-scale theory that provides an explicit model realisation of dimension seven operator mentioned earlier in this section.

The large leptonic mixing can also be accommodated within the model, as it possesses sufficient freedom and structurally distinct mass-generation mechanisms for charged fermions and neutrinos. Note that the charged-lepton mass matrix is not diagonal in the aforementioned basis, and therefore a non-trivial left-handed rotation matrix $u_{eL}$ already arises. Moreover, the flavour structure of the light neutrino mass matrix $M_\nu$ can be tailored to the desired form by choosing an appropriate structure for the $\mu$ matrix. Indeed, the apparent difference in the origin of the charged-fermion and neutrino mass matrices, as reflected in Eqs. (\ref{mf_eff},\ref{M_nu}), suggests that leptonic mixing is structurally distinct from quark mixing. A detailed construction of neutrino mass models, lepton-mixing patterns, and their phenomenological implications is left for future work, as the primary focus of the present study is the charged-fermion sector.

\section{Summary and Discussions}
\label{sec:summary}
We have presented an explicit realization of the recently proposed mechanism for generating hierarchical masses among multiple copies of fields, applied here to the flavour structure of the SM. A minimal and phenomenologically viable implementation requires five pairs of weak isosinglet vectorlike fermions for each of the up-type quark, down-type quark, and charged lepton sectors. The nearest-neighbour mass terms form a one-dimensional lattice in theory space, and specific attachments of the SM chiral fermions ensure that only a single generation acquires mass at tree level. The masses of the first- and second-generation fermions then emerge from non-local effects induced by radiative corrections along the lattice. The mechanism is distinct from other lattice-based theory-space constructions, both in its origin of flavour hierarchies and in its phenomenological consequences, as explored and described in this work.

The radiative corrections are incorporated through an abelian gauge extension under which the SM fermions remain neutral, while the vectorlike fermions are universally charged. Nevertheless, flavour-violating vertices involving the SM fermions and the new gauge boson, as well as the $Z$ and Higgs bosons, arise through heavy–light fermion mixing. The masses of the five copies of vectorlike fermions are necessarily non-degenerate and are reasonably spread around the new-physics scale. We perform a detailed numerical analysis and obtain several benchmark solutions to study the resulting mass spectra and couplings within the underlying framework. We also examine the phenomenological implications, covering a broad set of the most relevant observables.

It is observed that, while the mechanism can be effectively and consistently realised at any UV scale above the electroweak scale, a realisation closer to the electroweak scale is preferred from the perspective of fine-tuning required to maintain a separation between the two scales. At the same time, the new-physics scale cannot be too close to the electroweak scale in view of existing constraints from various direct and indirect searches for new physics, which apply effectively to this model. Taking all effects into account, we find that a relatively low-scale realisation of this mechanism is feasible, with the lightest vectorlike fermions and the new gauge boson having masses around $5$ TeV.

Let us specifically compare this mechanism and its phenomenological aspects with the conventional radiative mass-generation mechanisms discussed recently in \cite{Weinberg:2020zba,Jana:2021tlx,Mohanta:2022seo,Mohanta:2023soi,Mohanta:2024wcr,Jana:2024icm,Mohanta:2024wmh,Mohanta:2025wiq}.
\begin{itemize}
\item The new physics required for radiative mass generation need not involve ${\cal O}(1)$ flavourful couplings. In the present framework, all such couplings arise only at ${\cal O}(\rho^{2})$, where $\rho$ typically denotes the ratio of SM to vectorlike fermion masses. This allows a substantial reduction of the new-physics scale. As demonstrated in this work, constraints from various sectors still permit new fermions and the gauge boson with masses as low as $5$ TeV, compared to the lower bound of $\sim 200$ TeV obtained in the most optimised scenarios of conventional radiative mechanisms.
\item The scalar sector of the present framework is particularly simple, with the SM Higgs extended by only one complex scalar. Models in other class typically require multiple Higgs doublets, and only in a specific alignment limit \cite{Mohanta:2025wiq} do the Higgs couplings to fermions become SM-like. As shown in Section~\ref{subsec:BSMH}, the Higgs couplings to fermions in the present model are automatically proportional to the fermion masses, with deviations arising only at ${\cal O}(\rho^{2})$.
\item In conventional approaches, a single pair of vectorlike fermions per sector is sufficient to generate rank-one tree-level mass matrices, whereas in the present framework a minimum of five pairs is required. Despite this, the number of parameters introduced in the latter case remains modest due to the nearest-neighbour structure of the interactions among them.
\end{itemize}
The framework presented here is also qualitatively distinct from the universal seesaw mechanism \cite{Berezhiani:1983hm,Davidson:1987mh}, in which the masses of all generations are generated at tree level through the addition of three vectorlike families with generic couplings.

While a substantial reduction of the new-physics scale is achieved compared to models based on the conventional radiative mass-generation mechanism, the number of free parameters remains comparable. Further improvement may be attained by implementing the aforementioned mechanism in theories that unify quarks and leptons and/or by embedding chiral and vectorlike flavours into irreducible representations of a flavour group. The latter could replace the $U(1)_X$ symmetry with a non-abelian alternative, which is also desirable from the perspective of maintaining perturbative validity beyond $M_X$, as discussed earlier. Different configurations of lattice structure may also be explored, in particular to decouple the top quark mass from being generated through the seesaw mechanism. A structure capable of explaining the hierarchy, $m_{b, \tau} \ll m_t$, would also be desired. These considerations may be explored further in future works.

\begin{acknowledgments}
GM acknowledges the support by an appointment to the JRG Program at the APCTP through the Science and Technology Promotion Fund and Lottery Fund of the Korean Government and by the Korean Local Governments – Gyeongsangbukdo Province and Pohang City. GM also acknowledges the support from Physical Research Laboratory where initial part of this work was carried out. The work of KMP is supported by the Department of Space (DOS), Government of India.
\end{acknowledgments}

\appendix
\section{Magnitude of BSM couplings}
\label{app:omega}
In this appendix, we list the absolute values of the couplings $\Omega_{fL,fR}$ defined in Eq. (\ref{omega}) for the three benchmark solutions S1, S2 and S3. They quantify couplings of the SM fermions with $X$ boson as well as the deviation in the couplings with $Z$ and Higgs bosons as discussed in Section \ref{sec:bsm}. The numerically computed values are listed in Table~\ref{tab:Sigma}. 
\begin{table}[t]
\centering
\setlength{\tabcolsep}{6pt}
\renewcommand{\arraystretch}{1.00}
\begin{tabular}{c c c c}
\toprule
Solutions & $f$ & $|\Omega_{fL}|$ & $|\Omega_{fR}|$ \\
\midrule
\multirow{7}{*}{S1}
& $u$ &
$10^{-2}\!
\begin{pmatrix}
0.15 & 0.19 & 0.11\\
0.19 & 4.3  & 1.1 \\
0.11 & 1.1  & 4.5
\end{pmatrix}$
&
$\begin{pmatrix}
3.5\times10^{-6} & 3.5\times10^{-4} & 1.1\times10^{-3} \\
3.5\times10^{-4} & 0.042 & 0.089 \\
1.1\times10^{-3} & 0.089 & 0.80
\end{pmatrix}$
\\

& $d$  &
$10^{-4}\!
\begin{pmatrix}
0.083 & 0.19 & 0.042\\
0.19  & 0.80 & 0.15 \\
0.042 & 0.15 & 0.84
\end{pmatrix}$
&
$10^{-3}\!
\begin{pmatrix}
0.14 & 0.30 & 0.11\\
0.30 & 1.1  & 0.31\\
0.11 & 0.31 & 1.3
\end{pmatrix}$
\\

& $e$ &
$10^{-4}\!
\begin{pmatrix}
0.027 & 0.21 & 0.079\\
0.21  & 2.3  & 0.82 \\
0.079 & 0.82 & 3.0
\end{pmatrix}$
&
$10^{-3}\!
\begin{pmatrix}
0.047 & 0.37 & 0.15\\
0.37  & 4.1  & 1.5 \\
0.15  & 1.5  & 5.3
\end{pmatrix}$
\\
\midrule

\multirow{7}{*}{S2}
& $u$ &
$10^{-3}\!
\begin{pmatrix}
0.15 & 0.092 & 0.079\\
0.092 & 2.6 & 0.60\\
0.079 & 0.60 & 2.6
\end{pmatrix}$
&
$\begin{pmatrix}
4.6\times10^{-6} & 5.6\times10^{-4} & 1.2\times10^{-3} \\
5.6\times10^{-4} & 0.083 & 0.13 \\
1.2\times10^{-3} & 0.13 & 0.81
\end{pmatrix}$
\\

& $d$ &
$10^{-5}\!
\begin{pmatrix}
0.13 & 0.25 & 0.064\\
0.25 & 1.0  & 0.22 \\
0.064& 0.22 & 1.1
\end{pmatrix}$
&
$10^{-4}\!
\begin{pmatrix}
0.48 & 1.0 & 0.32\\
1.0  & 4.6 & 1.3 \\
0.32 & 1.3 & 4.9
\end{pmatrix}$
\\

& $e$ &
$10^{-5}\!
\begin{pmatrix}
0.0090 & 0.078 & 0.030\\
0.078  & 0.93  & 0.34 \\
0.030  & 0.34  & 1.2
\end{pmatrix}$
&
$10^{-2}\!
\begin{pmatrix}
0.012 & 0.089 & 0.053\\
0.089 & 0.94  & 0.53 \\
0.053 & 0.53  & 1.6
\end{pmatrix}$
\\
\midrule

\multirow{7}{*}{ S3}
& $u$ &
$10^{-4}\!
\begin{pmatrix}
0.31 & 0.29 & 0.20\\
0.29 & 5.8  & 1.8 \\
0.20 & 1.8  & 6.4
\end{pmatrix}$
&
$\begin{pmatrix}
3.5\times10^{-6} & 3.9\times10^{-4} & 8.9\times10^{-4} \\
3.9\times10^{-4} & 0.052 & 0.084 \\
8.9\times10^{-4} & 0.084 & 0.57
\end{pmatrix}$
\\

& $d$ &
$10^{-7}\!
\begin{pmatrix}
0.27 & 0.56 & 0.22\\
0.56 & 2.4  & 0.86\\
0.22 & 0.86 & 2.9
\end{pmatrix}$
&
$10^{-3}\!
\begin{pmatrix}
0.077 & 0.14 & 0.13\\
0.14  & 0.46 & 0.34\\
0.13  & 0.34 & 0.91
\end{pmatrix}$
\\

& $e$ &
$10^{-6}\!
\begin{pmatrix}
0.09 & 0.63 & 0.30\\
0.63 & 6.6  & 2.8 \\
0.30 & 2.8  & 9.5
\end{pmatrix}$
&
$10^{-3}\!
\begin{pmatrix}
0.036 & 0.27 & 0.10\\
0.27  & 2.9  & 1.0 \\
0.10  & 1.0  & 3.9
\end{pmatrix}$
\\[6pt]
\bottomrule
\end{tabular}
\caption{Absolute values of the matrices $\Omega_{fL}$ and $\Omega_{fR}$ for $f=u,d,e$ and for all three benchmark solutions.}
\label{tab:Sigma}
\end{table}

\section{Form factors for $f_1 \to f_2 \gamma$ computation}
\label{app:Formfactors}
The form factors used in the computation of $\Gamma(t \to \gamma\,q_a)$ and $\Gamma(t \to g\,q_a)$ in Section \ref{subsec:rare_top} are given in this appendix. The same can be modified appropriately for the computation of the partial widths of radiative decays of the charged leptons. 

The explicit expressions of $A^\gamma_{q_a}$ and $B^\gamma_{q_a}$ appearing in Eq. (\ref{MA:tcg}) are given as \cite{Aguilar-Saavedra:2002lwv}
\beqa \label{FF:A}
A^\gamma_{q_a} &=& \sum_\alpha \frac{g^2 c_\theta^2}{c_W^2} \frac{Q_\alpha e}{16 \pi^2}\,\Big[c^{LL}_{a \alpha} \Big\{-(m_t+m_{q_a}) C_0  - (2m_t+m_{q_a}) C_1 - (m_t + 2 m_{q_a}) C_2\, \nonumber \\
&-& m_t C_{11} - (m_t + m_{q_a}) C_{12} - m_{q_a} C_{22}\Big\} + \left( c^{LR}_{a \alpha} + c^{RL}_{a \alpha}\right) 2 \overline{m}_{u_\alpha} (C_0+C_1+C_2) \Big]\,, \nonumber \\
B^\gamma_{q_a} &=& \sum_\alpha \sum_\alpha \frac{g^2 c_\theta^2}{c_W^2} \frac{Q_\alpha e}{16 \pi^2}\,\Big[c^{LL}_{a \alpha} \Big\{-(m_t-m_{q_a}) C_0  - (2m_t-m_{q_a}) C_1 - (m_t - 2 m_{q_a}) C_2\, \nonumber \\
&-& m_t C_{11} - (m_t - m_{q_a}) C_{12} + m_{q_a} C_{22}\Big\} + \left( c^{LR}_{a \alpha} - c^{RL}_{a \alpha}\right) 2 \overline{m}_{u_\alpha} (C_0+C_1+C_2) \Big].
\eeqa
Here, $m_t$ and $m_{q_a}$ are the masses of external quarks. ${\overline{m}_{u_\alpha}}$ are the masses of internal quarks in $\overline{\rm MS}$ renormalisation scheme. $Q_\alpha$ is an electromagnetic charge of $\alpha$ flavour of fermions propagating in the loop. Various $C_{...}$ are loop integration factors involving the internal and external masses with arguments $C_{...} \equiv C_{...}[m_t^2,0,m_{q_a}^2,M_Z^2,\overline{m}_{u_\alpha}^2,\overline{m}_{u_\alpha}^2]$. Their explicit expressions are given in \cite{Denner:1991kt}. In the present work, we numerically compute them using \textsc{LoopTools} \cite{Hahn:1998yk}. The expressions for $A^g_{q_a}$ and $B^g_{q_a}$ can be obtained by replacing $e$ with $g_3$ and setting $Q_\alpha = 1$ in the above expressions.

The coefficients $c^{...}_{a\alpha}$ parametrise the new physics contributions. They are computed for the present model as,
\beqa \label{FF:coeff}
c^{LL}_{a\alpha} &=& \left[ \left(T^u_3- Q_u s_W^2  \right)\,\delta_{a \alpha} + \left(\delta g^Z_{uL}\right)_{a \alpha} \right] \left[ \left(T^u_3- Q_u s_W^2 \right)\,\delta_{3 \alpha} + \left(\delta g^Z_{uL}\right)_{3 \alpha} \right]\,,\nonumber \\
c^{LR}_{a\alpha} &=& \left[ \left(T^u_3- Q_u s_W^2  \right)\,\delta_{a \alpha} + \left(\delta g^Z_{uL}\right)_{a \alpha} \right] \left[\left(-Q_u s_W^2  \right)\,\delta_{3 \alpha} \right]\,,\nonumber \\
c^{RL}_{a\alpha} &=&\left[\left(- Q_u s_W^2  \right)\,\delta_{a \alpha} \right]  \left[\left(T^u_3- Q_u s_W^2  \right)\,\delta_{3 \alpha} + \left(\delta g^Z_{uL}\right)_{3 \alpha} \right]\,.
\eeqa
It is straightforward to see that all the $c^{...}_{a\alpha}$ vanish for $a=1,2$ if the off-diagonal elements in $\delta g_{uL}^Z$ are zero. The similar expressions hold for the charged-leptons, with appropriate replacement of fermion indices in Eqs. (\ref{FF:A},\ref{FF:coeff}).

\bibliography{biblio}
\bibliographystyle{JHEP.bst}
\end{document}